# Polarisation optics for biomedical and clinical applications: a review

Chao He[1,*], Honghui He[2,3,*], Jintao Chang[2,4], Binguo Chen[2,5], Hui Ma[2,3,4] and Martin J. Booth[1,*]

[1]*Department of Engineering Science, University of Oxford, Parks Road, Oxford, OX1 3PJ, UK*

[2]*Guangdong Engineering Center of Polarisation Imaging and Sensing Technology, Tsinghua Shenzhen International Graduate School, Tsinghua University, Shenzhen 518055, China*

[3]*Institute of Biopharmaceutical and Health Engineering, Tsinghua Shenzhen International Graduate School, Tsinghua University, Shenzhen 518055, China*

[4]*Department of Physics, Tsinghua University, Beijing 100084, China*

[5]*Department of Biomedical Engineering, Tsinghua University, Beijing 100084, China*

[*]*Corresponding authors: chao.he@eng.ox.ac.uk; he.honghui@sz.tsinghua.edu.cn; martin.booth@eng.ox.ac.uk*

**Many polarisation techniques have been harnessed for decades in biological and clinical research, each based upon measurement of the vectorial properties of light or the vectorial transformations imposed on light by objects. Various advanced vector measurement/sensing techniques, physical interpretation methods, and approaches to analyse biomedically relevant information have been developed and harnessed. In this review, we focus mainly on summarizing methodologies and applications related to tissue polarimetry, with an emphasis on the adoption of the Stokes-Mueller formalism. Several recent breakthroughs, development trends, and potential multi-modal uses in conjunction with other techniques are also presented. The primary goal of the review is to give the reader a general overview in the use of vectorial information that can be obtained by polarisation optics for applications in biomedical and clinical research.**

**Introduction**

Light, as an electromagnetic wave, possesses several fundamental properties, which include intensity, wavelength, phase and polarisation[1,2] (see Fig. 1a). While the former three are scalar quantities, polarisation has vectorial properties; its use has therefore required more advanced optical and computational approaches. Hence, studies of either the vector properties of light, described via the state of polarisation (SOP) or the full vectorial transformation properties of an object, have a shorter history in biomedical analysis compared with their scalar counterparts, and the extent of their application is still being explored[3-5]. So far, numerous intriguing areas of research have been enhanced through harnessing vectorial information acquired via polarisation optics; these range from fundamental research[6-10], such as quantified polarisation entropy[11], across quantum physics[12], such as spin-orbital interaction of light[13,14], to material characterisation (e.g. chiral characteristics[15]) or for biomedical studies and clinical applications (e.g. characterisation of structural features in tissue[16-21]).

Scattering, especially through multiple scattering processes, alters the degree of polarisation and SOP of the incident light beam[22]. While it is an insightful procedure for evaluating structural information of biomedical samples including tissues and cells[16], it also introduces uncertainty in expected photon properties[22]. This characteristic largely hinders the development of modern tissue polarimetric techniques and related information analysis[20,22,23]. The turbidity of many tissue structures imposes randomness on the photons' interaction processes, which complicates the detection and analysis of vectorial information[20]. Such phenomena also distinguish tissue polarimetry from the traditional polarisation measurement technique of ellipsometry[22-26]. As summarized in Fig. 1b, their comparison shows several commonalities and differences. The Jones formalism is used for clear and non-depolarising media such as thin films; it consists of the Jones vector (describing the polarisation property of the light) and Jones matrix (describing the polarisation transformation properties of the object). They have been widely used in ellipsometry techniques[25,26] (see Fig. 1b; and summary in Ref[26]). Another polarisation formalism is Stokes-Mueller, in which the Stokes vector and the Mueller matrix are used to describe the light beam and the object, respectively. Neither the Stokes vector nor the Mueller matrix maintain absolute phase information, but have the advantage of being able to represent depolarisation[27,28]. This is often essential in biomedical polarimetry, whose applications normally involve scattering induced light depolarisation[20-23]. There exists an increasing trend in both modern ellipsometry and polarimetry to deal with increasingly complex media, moving from isotropic and homogeneous media towards anisotropic and inhomogeneous ones[20-31]. While modern ellipsometry is developing towards full polarisation measurement using the Stokes-Mueller formalism, advanced polarimetry is gradually changing from full vectorial measurement to partial detection, as some key features of biomedical specimens could possibly be revealed through partial, rather than complete, measurements of vectorial information[16,32,33].

The structure of this review is given in Fig. 1c; it consists of introducing the basic polarisation optical tools, summarizing the current vectorial information detection, extraction, and analysis approaches, and pointing out the possibilities for future multi-modal synergy with other cutting-edge technologies. Although biomedical polarimetry is still developing towards various research fields and applications, largely unexplored spaces still exist. We also hope this review could stimulate new explorations or breakthroughs in such prospective fields.

It is worth noting that the use of biomedical polarimetry is expanding and has also been summarized in several recent reviews by **Tuchin**[22], **Ghosh & Vitkin**[20], **Ramella-Roman** et al.[23], **Qi & Elson**[24], **De Boer** et al.[34], **He** et al.[16]. They have demonstrated the fast progress of this technique in the biomedical and clinical fields.

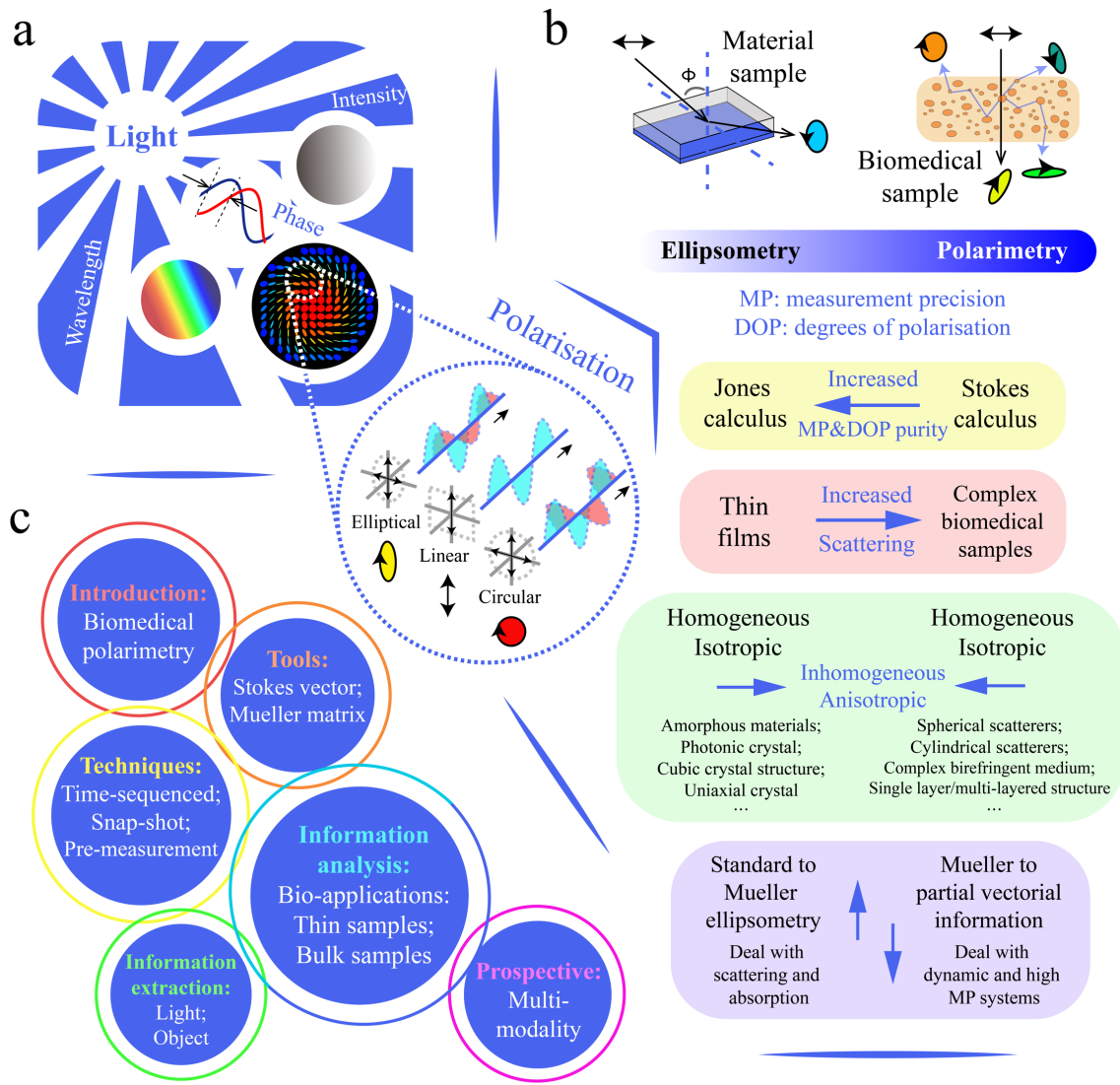

**Figure 1: Developing trends of tissue polarimetry and the structure of this review.** (a) Light properties: intensity, wavelength, phase and polarisation. (b) The comparisons of the development of modern ellipsometry and polarimetry. The inset blue arrows in different colour boxes represent either changing directions or developing trends. (c) The structure of this review.

## 1 Fundamental vectorial representation for polarised optics in biomedical applications

Sample induced scattering is prevalent in biomedical imaging, particularly in tissues[16,20,22,23]. This introduces additional SOP modulations that affect diattenuation and retardance as well as depolarisation[16,22,23]. A complex scattering medium can often be modelled by several basic components, like spherical scatterers of different sizes[35]; cylindrical rod-like scatterers with different orientational distributions[36]; and birefringence for an interstitial medium[37-39]; combinations of these features can all be adjusted to mimic the real object[36,40-42]. Other physical conditions such as layer complexity (single-layered or multi-layered scattering[43]), or scattering type (elastic domain for Mie and Rayleigh scattering; or inelastic domain, like Raman scattering) are also described in the literature, e.g., see Ref[44]. Modelling of the scattering assumptions can be conducted via **Monte Carlo simulation**[45,46]. This is a widely used statistical method for quantitative analysis of the interactions between polarised photons and complex biomedical media[40-42], especially bulk media with multiple-scattering properties, for which the analytical solutions to describe the interactions cannot be obtained. In this review, we focus on the occurrence of elastic scattering in conjunction with other polarisation characteristics (see Fig. 2a) for biomedical polarimetry.

In the presence of depolarisation, Jones calculus, which represents only transitions between pure polarisation states, is of limited use as it cannot comprehensively describe the light properties, especially the degree of polarisation for partially polarised light[2-5]. Intrinsically, Jones calculus is based on the assumption that electric field vector holds a particular stationary state. For partially polarised light (or fully depolarised light), the variation of the electrical vector as the light propagates is semi-disordered (or completely disordered) so that more degrees of freedom are required to describe the light field[47]. In the scope of linear optics, the Stokes vector, which is a $4 \times 1$ vector, is used to characterise the SOP of the light beam[47,48]; while the Mueller matrix, which is a $4 \times 4$ matrix, describes the transformation properties of the object that affect the Stokes vector[47,48]. Hence, considering that the scope of this review focuses on tissue polarimetry, we place an emphasis on the Stokes-Mueller formalism.

*Stokes vector*

The Stokes vector can be expressed with the format shown in Fig. 2b[47,48]; where

$$I = I_0 + I_{90}$$

$$Q = I_0 - I_{90}$$

$$U = I_{45} - I_{-45}$$

$$V = I_R - I_L$$

$I_0$, $I_{90}$, $I_{45}$, $I_{-45}$ are the projection intensities (different linear components in directions of 0°, 90°, 45°, -45° with respect to the local coordinate system) of a light beam, $I_L$ and $I_R$ are components of left/right-handed circular polarised light, respectively. Note some other parameters can be defined with components of Stokes vector: degree of polarisation

$$DOP = \sqrt{Q^2 + U^2 + V^2}/I$$

degree of linear polarisation

$$DOLP = \sqrt{Q^2 + U^2}/I$$

and degree of circular polarisation of light[3-5].

$$DOCP = \sqrt{V^2}/I$$

From the above expressions, we see that the Stokes vector can be calculated via intensity measurements that can be readily performed in an experiment[47,48]. The Jones vector, on the other hand, is defined by amplitude and phase that cannot be directly measured, which is another reason why the Jones approach is less well suited to biomedical polarimetry[20-23]. The intrinsic reason for the existence of depolarisation is due to temporal or spatial averaging[16,20-23]. If an extremely fast and small detector could monitor the vector properties of the light, then it would only detect polarised light. Such averaging properties can also be found in the definition of the Stokes vector[47,48]. Note the definitions of right-handed circular polarised light (clockwise rotation) and left-handed circular polarised light (anticlockwise rotation) are different in optics books and academic communities. It depends on whether the observer 'sees' the light from the source (**Convention I**), or from the detector (**Convention II**). Institute of Electrical and Electronics Engineers (IEEE) uses **Convention I**, so it is also widely used in engineering fields; Quantum physicists also use **Convention I**, to be consistent with the conventions for representing particle spin states[49,50]. However, for numerous optics books such as Principles of Optics (Born & Wolf[48]) and Handbook of Optics[51], **Convention II** is used. In this review, we use **Convention II** in order to correspond to such scientific references.

The Jones vector has a graphical representation known as the polarisation ellipse[47,48] (if we add the parameter DOP, the polarisation ellipse can also represent the Stokes vector (see Fig. 2b (i))). While for Stokes vector visualization, the Poincaré sphere (PS) is commonly used[47,48] (see Fig. 2b (ii)). SOPs are represented via the PS, which is defined in a three-dimensional coordinate system, whose coordinates correspond to the eigenbasis formed by $Q$, $U$ and $V$ (each normalized by $I$). The PS is a unitary sphere that represents complete polarisation states on its surface and depolarised states inside the sphere. Any transformation of the SOP through a specimen is equivalent to manipulation of the original Stokes vector between different points on or inside the PS. Figure 2b (ii) gives a schematic demonstration of the PS. The length of the vector from the origin point to the SOP location denotes the DOP[47,48]. The letters $H, V, M, P$ are specific polarisation states: horizontally polarised ($H$), vertically polarised ($V$), 45° polarised ($M$) and -45° polarised ($P$). The polarisation ellipse parameters ($\chi$ and $\psi$) can be interpreted from the azimuth angle (and the polar angle) of the derived vector inside the PS.

Such a graphical representation excludes the absolute phase information, which is sometimes not addressed in typical vectorial beam analysis such as pure polarisation measurement or tissue polarimetry[20-23,47,48]. However, one type of the absolute phase variation that is referred to as geometric phase is related to the pathway of the SOP movement on the surface of the PS[47,48], which features scope for the extension of the tissue polarimetric technique (also see Discussion).

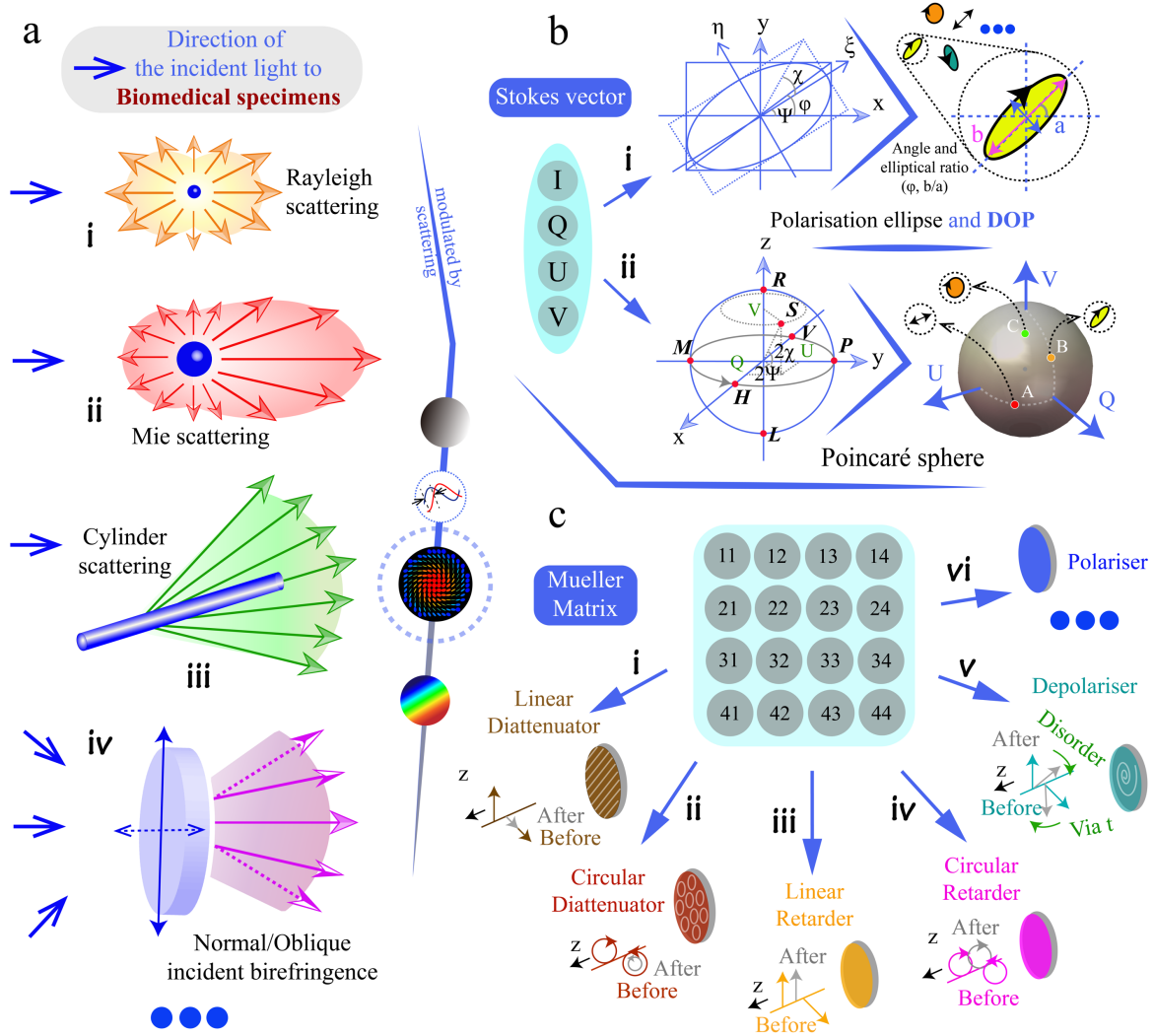

**Figure 2: Fundamental light-matter interaction processes in biomedical specimens and polarisation tools.** (a) Several scattering types and media manipulation, showing schematically the way in which light is scattered in the differing scenarios[16-24]. (b) The Stokes vector representations. Top row: the polarisation ellipse and the DOP can be used to represent a given Stokes vector. Bottom row: the Poincaré sphere visualizes all states of polarisation with linear states on the equator, circular states on the north/south poles and elliptical states in between[2-5]. (c) Vectorial properties that can be encoded in the Mueller matrix. Linear/circular diattenuator, linear/circular retarder etc. are fundamental polarisation elements; the arrows on the lines or circles represent the eigenbasis of modulated light beam passing through such media, where 'before' and 'after' illustrate the amplitude and/or relative phase change of the chosen eigenbasis. For further details refer to Ref[2-5].

*Mueller matrix*

The Mueller matrix (MM) describes the vectorial transformation properties of an object[16,20-23]. As illustrated in Fig. 2c, the MM describes the transformation of one Stokes vector into another. The MM represents the full vector properties of an object through its 16 elements ($m_{kl}; k, l = 1,2,3,4$). Among these, $m_{11}$ represents the transformation of scalar intensity (absorption or other loss); the other 15 elements encode the vectorial properties of the object[47,48]. Direct physical meanings of these 15 elements taken individually are normally ambiguous[16,20-23]. As illustrated in Fig. 2c, several fundamental polarisation properties are encoded in (and can be extracted from) the MM. They are linear/circular diattenuation, linear/circular retardance, linear/circular polarisance, linear/circular depolarisation and so on[2-5,52-62]. The effects of each of these fundamental optical mechanisms on the light vectors along the propagation direction z are shown in Fig 2c (i) to (vi), where 'before' and 'after' illustrate the amplitude and/or relative phase change of the chosen eigenbasis. Amongst these mechanisms, the diattenuator possesses two different absorption ratios for two polarisation directions; it in effect reduces the intensity of one polarisation compared to the other. The retarder exhibits different refractive indices for two polarised eigenvectors, in effect leading to an additional relative phase difference between the two vectors. The depolariser can modify the DOP of the light beams. For more detailed descriptions of the mechanisms and further examples see Ref[2-5].

Both Stokes vectors and MMs can represent the effects of time-averaged induced depolarisation[16]. An object may introduce two different classes of depolarisation: homogenous depolarisation and inhomogeneous depolarisation. The former one can lead to a similar DOP change for any SOP; such properties can be observed in media such as a polystyrene sphere solution. The later one can lead to different DOP change for different SOP; typical examples are found in complex biomedical tissue.

Several factors may contribute to depolarisation in experimental scenarios. We describe **three** main reasons here. a) The first reason relates to the **time domain.** In general, the Stokes vector polarimeter is based on intensity measurement[26], so in practice the intensity recorded at the detector includes a time-integration process. If the SOP changes rapidly, possibly due to multi-scattering induced by complex bio-media, then depolarisation would be measured. b) This reason relates to the **spatial domain.** When imaging processes are involved, every point on the beam section is created through the integration of various sub-beams that could have different polarisation states. The superposition of these states leads to depolarisation. c) The final reason is given in the **spectral domain.** Many processes that affect polarisation, such as birefringence and scattering, are also dependent on wavelength. Hence for different wavelengths, variations in amplitude and phase may also lead to depolarisation.

## 2 Vectorial information measurement techniques for biomedical applications

Numerous vectorial information measurement methods have been put forward in the past decades[4,7,11,26,28,29,63]. In this section, we categorize the polarisation measurement techniques into two types: time-sequenced and snap-shot approaches[28,29,64-67] (see Fig. 3). For both cases, the preparation required before detection is similar and can be divided into three general steps: denoising, optimization and calibration[32,68-78] (see Fig. 4). The aim of those steps is to reduce the complex errors that would occur during the measurement process, hence obtaining imaging results with higher precision and accuracy[32,69-72]. The technical aspects of such advanced polarimetry are summarized in the review papers by **Azzam**[28], **Chipman**[63] and **Tyo**[79].

Before full Stokes vector/MM measurements became widely adopted, there was successful work using fixed input SOP and fixed analyzers to perform partial vectorial detection for biomedical applications. **Jacques** et al. showed crossed-polarised light imaging to enhance surface contrast, detect skin cancer and other lesion margins[80,81]; **Demos** et al. added the dimension of wavelength based on crossed SOPs[82,83]; **Groner** et al. noted such techniques can enhance superficial vascular contrast, and hence adopt it into brain perfusion, pancreatic and further clinical diagnoses[84]; **Bargo** et al. took angle-dependency into consideration when measuring skin tissue[85]. **Sridhar** et al. also studied multiply scattered photons to enhance information extraction from biological specimens via elliptically polarised light[86].

Both time-sequenced and snap-shot polarimetry techniques can be classified in two general ways: firstly, as either Stokes vector (light property) or MM (material property) measurement; and secondly, as partial or full vectorial measurement (Fig. 3). We will classify different techniques using the second criteria in later sections of this review.

*Time-sequenced techniques*

Stokes polarimetry is clearly the basis for more advanced MM polarimetry. Both of their intrinsic mechanisms can be interpreted with respect to the instrument matrix (*A*)[68-72] (see Fig. 4). This matrix represents the settings of the polarisation state generator (PSG) and polarisation state analyser (PSA) in the various measurement steps: for MM measurement it represents the PSG and PSA, for Stokes vector measurement it represents one the PSA. Combinations of the rotating waveplate and/or polariser are widely adopted in such approaches[64-66]. The original proposal for a Stokes vector measurement scheme (that using SOPs of *H*, *V*, *M*, *P, L* and *R*) was from **Collett**[87] in 1984. Later it was adopted for biomedical information extraction or phantom analysis with ability of the full depolarisation information characterisation[16,20-23].

The use of rotating components has disadvantages, such as increasing measurement time and introducing unexpected errors from mechanical movements. However, such systems are easy to construct. Hence, numerous commercialized polarimeters still use this approach. In order to make improvements, researchers have tried to reduce the number of the rotating components (such as the dual-rotating waveplate MM polarimeter with fixed polarisers that was proposed by **Azzam**[64] in 1978, which is widely used in tissue analysis[16,22]) or use fast modulation devices (such as Stokes or MM polarimeters enabled via liquid crystal variable retarders (LCVR)[88], spatial light modulators (SLM)[89], ferroelectric liquid crystals (FLC)[90], or photoelastic modulators (PEM)[91]). Besides full MM detection, partial MM measurement, such as 3 × 3 MM imaging of linear polarisation states, also gained wide attention. **Qi** et al. used related methods in analysing linear depolarisation and retardance of rat tissue[92].

Although there are some applications that require high speed operation, such as detection in dynamic situations like *in vivo* sensing for clinical diagnosis[24], time-sequenced polarimeters still play an important role in modern polarimetric research, due to their mature state of development and simple configuration. Such applications include characterisation of complex vector fields[6,7,9], or providing ground truth validation in tissue research (e.g., differentiating human breast cancer[93-95]).

*Snap-shot techniques*

Rapidly changing or dynamic objects need snap-shot detection, in order to correctly extract vectorial information that would be complicated by time-sequenced measurement. Snap-shot approaches are configured to take different measurements in parallel, as opposed to the serial measurement of sequential techniques. In general, snap-shot techniques must, to some degree, sacrifice alternative dimensions to enable simultaneous vector measurement[67,96]. Those methods include (see Fig. 3b): Stokes vector polarimeters with division-of-amplitude[97-99], division-of-wavefront[71,100-105] or division-of-focus-plane[106-122] – these fit in the category of spatial modulation with respect to different analysis channels (see Fig. 3b (iii)). Savart-plate-based polarimeters (**Oka** et al.) are in the category of Fourier frequency domain segmentation, which are interferometric systems where the polarisation information is encoded in the spatial carrier fringes[123]. If combined with the property of birefringence dispersion, spectroscopic polarimetry with channelled spectrum can also be presented[67,96].

Similar to Stokes vector polarimeters, there exist concepts for snap-shot MM polarimeters, in which certain dimensions are sacrificed to enable simultaneous MM estimation (see 3b (iv)). **Dubreuil** et al.[67] and **Hagen** et al.[96] utilized different wavelength-dependent birefringent media to resolve the MM in a single shot, within the limitation of the sample being achromatic. **Piquero** et al.[124] utilized full Poincaré beams as a PSG, enabling MM polarimetry with division of the wavefront. **He** et al. applied a spatially segmented method with defocusing to measure statistically averaged properties of biomedical samples[72]. As complete snap-shot MM techniques are rather complex, their usage for extractions of biomedical information is less common than the use of single-shot partial MM or Stokes vector polarimetry. For instance, $3 \times 4$ MM imaging is also gaining attention using circularly polarised illumination[32]; **Chang** et al. brought such a technique into human liver and cervical carcinoma tissue analysis[32] (see 3b (ii)).

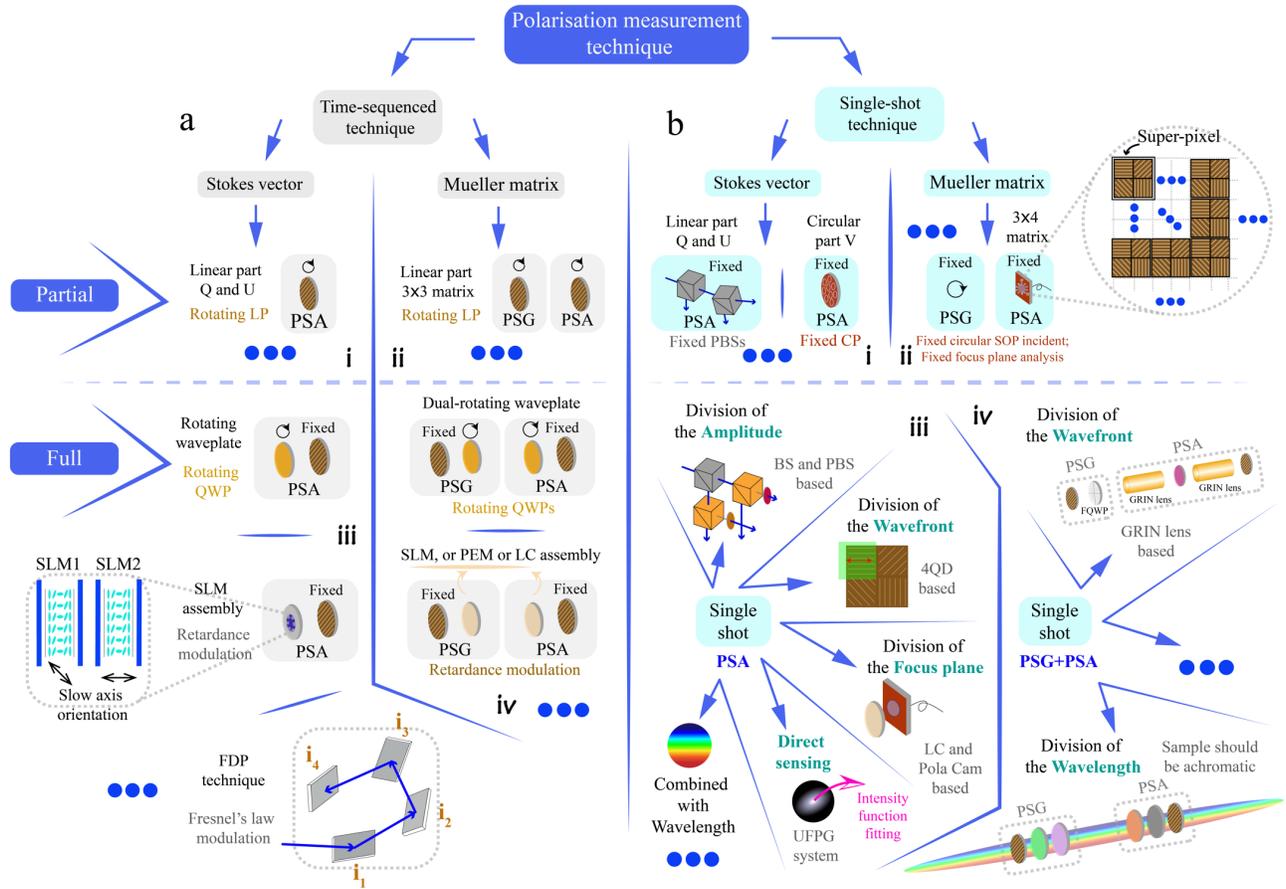

**Figure 3 Time sequenced and single shot polarisation measurement techniques.** (a) Time sequenced techniques: (i) Partial Stokes vector polarimetry; the PSA can be a rotating polariser. (ii) Partial MM polarimetry; the PSG and PSA can both be a rotating polariser. (iii) Full Stokes vector polarimetry; the PSA and PSG can both be a tuneable retarder (rotating quarter wave plate or SLM assembly) followed by a fixed polariser[28,63,79]. (iv) Full MM polarimetry; the PSG can be a fixed polariser followed by a tuneable retarder[28,63,79] (rotating quarter wave plate or LC components); the PSA can be a tuneable retarder (rotating quarter wave plate or LC components) followed by a fixed polariser[28,63,79]. (b) Snap-shot techniques: (i) Partial Stokes vector polarimetry; the PSA can be a fixed PBS assembly or circular polariser. (ii) Partial MM polarimetry; the PSA can be a fixed polariser array[32]. (iii) Full Stokes vector polarimetry; the types consist of division of the amplitude[97-99], division of the wavefront[71,100-105], division of the focus plane[106-122], and so on. (iv) Full MM polarimetry; the types consist of division of wavefront[72], division of the wavelength[67,96], and so on. (LP: linear polariser; QWP: quarter waveplate; FDP: four-detector photopolarimeter; for more information about such Fresnel's law based polarimetry refer to Ref[125]; BS: beam splitter; PBS: polarisation beam splitter; CP: circular polariser; FQD: four quadrant detector; FQWP: four quadrant wave plate; UFPG: universal full Poincaré generator)

*Denoising, optimization and calibration*

The measurement precision and sensitivity are vital in polarimetric techniques, hence the errors need to be properly controlled[28,32,68-78,126]. However, as Stokes vector or MM measurements belong to high dimensional information detection with multiple components[64,66,127], the error sources and error transfer process (such as accumulated amplifications through matrix calculations) are very complicated. Several previous analyses can be found in Ref [28,29,32,68-78,126-129]. In Fig. 4, we summarize in one diagram an overview of the structure of the 'denoising, optimization, and calibration' processes in polarimetric techniques with respect to random errors ($\delta\mathbf{A}$, $\delta\mathbf{I}$) and systematic errors ($\Delta\mathbf{A}$, $\Delta\mathbf{I}$). It also can be seen in the figure that three directions towards obtaining the correct vectorial measurements are still developing.

Note again that $\mathbf{A}$ is the instrument matrix for polarimetric measurement specifically, which is determined by the systematic configurations of polarisation optics and determines the error propagation amplification[20-24,69-72]; while $\mathbf{I}$ refers to the recorded intensity information. In Fig. 4, we take Stokes vector measurement equation ($\mathbf{S}=inv(\mathbf{A})\cdot\mathbf{I}$)[20-24,69-72] as a main illustration, to show the relationships between three steps in a picture, for simplicity. A similar structure (using the generalized equation: $\mathbf{M}=inv(\mathbf{A}')\cdot\mathbf{I}$) can be derived for the MM measurement, which is based fundamentally on the Stokes vector measurement process.

In order to reduce $\delta\mathbf{A}$ and $\delta\mathbf{I}$[32,68-78], a 'denoising process' is adopted. Figure 4 shows the approaches in time or spatial domain including time average and interpretation methods. To deal with the $\Delta\mathbf{A}$ and $\Delta\mathbf{I}$, a 'calibration process' is required. Numerous polarimetric calibration methods have been proposed[23,29,63,71,130]; these can be divided into global and local calibration approaches. Note that the calibration process itself also suffers from the error transfer process. Hence, determining the SOPs for calibration, choosing the standard calibration samples, as well as designing specific calibration methods for different systems should be taken into consideration[131,132]. Figure 4 also shows the process of 'optimization', which can deal with both types of errors, through global and local optimization approaches. For this process, different evaluation standards have been put forward to estimate the systematic performance. **Marenko** et al. considered the condition number (CN) in polarimetric optimisation[133], **Ambirajan, Tyo** and others have analyzed CN-based optimization on different phantoms[70,134-138]; and **Sabatke** et al. introduced equally weighted variance (EWV)[69] into the polarimetric area; **Azzam** et al. and following researchers explained the usage of geometry optimization based on Poincaré sphere internal volume (PSIV)[139-141]. Other useful criteria have also been proposed[142,143]. Such optimisation parameters can be used for evaluating the intrinsic error amplification of a polarimetry, which affect the accuracy and precision of the measurement[23,29,63,71,130-132]. If we consider the CN, the minimum CN value for a matrix based Stokes polarimetry is $\sqrt{3}$, which is the theoretical limit for systematic error amplification[68,70], as opposed to the minimum possible CN value (CN = 1) for matrix inversion. A similar error amplification also exists in MM polarimetry[144]. The three above-mentioned processes (denoising, optimization, calibration) are vital for any biomedical polarimetry, as they determine the credibility of the information extraction and further analysis.

For the matrix based calculation of Stokes polarimetry (within the scope of above explanations), there exist two problems: first, the mathematical aspect of minimal error amplification through the matrix calculation; second, the practical aspect that the above-mentioned three separate procedures contribute to error accumulation separately, as they require different evaluation criteria and are normally based upon different assumptions. In fact, there exists the possibility to jump out of the domain of matrix calculation for Stokes polarimetry, circumventing those drawbacks. An interesting direction is the adoption of a full Poincaré beam, taking advantage of its feature that maps all SOPs in a single beam[145]. **Vella**, **Zimmerman**, **He** and others have proposed different measurement approaches harnessing such beams based on different phantoms such as stress engineered optics[146-152] and graded index optics[68]. The full Poincaré beam Stokes vector technique has recently made it possible to have a clear information-based learning approach (such as the task of searching for the brightest points), combining the 'end to end' solution (a combination of above three processes – denoising, optimization and calibration) together for an enhanced polarimetric measurement precision and accuracy[68]. In essence, this approach means that the Stokes vector retrieval process changes from matrix-based calculation to information-based image processing.

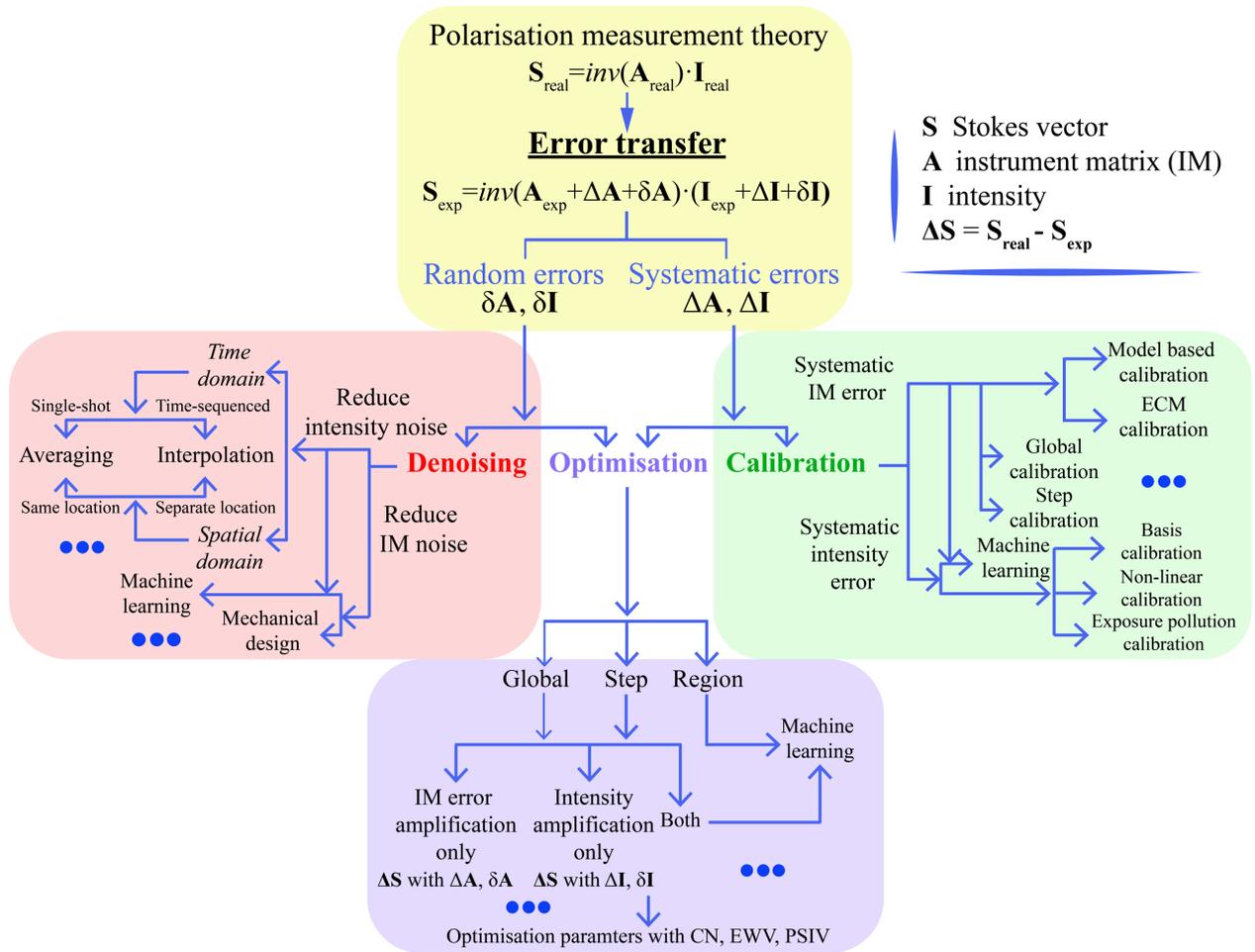

**Figure 4 Polarisation measurement theory for Stokes vectors.** The polarisation measurement theory is summarized in three aspects: denoising, optimisation, and calibration.

## 3 Vectorial information extraction methods for biomedical applications

Information about the vectorial properties of a biological specimen can be derived partially from the polarisation properties of the light beam or, in a more complete fashion, from the polarisation properties of the tissue itself [20-23,153]. To extract information from the measured Stokes vector or MM (or part of them), different decomposition methods and parameters were proposed to represent meaningful physical processes, to extract information that could be used in subsequent analysis[52-62].

*Information extraction from the vector properties of the light beam*

Several parameters can be calculated from the Stokes vector directly (see Fig. 5a (i) and previous section): such as the degree of polarisation (DOP), degree of linear polarisation (DOLP) and degree of circular polarisation (DOCP) of light. For a single uniform light beam, the DOP is 1 for fully polarised, 0 for unpolarised or completely depolarised, and between 0 and 1 for partially polarised. The DOP cannot be larger than 1. Despite containing four elements, a Stokes vector contains fewer than four degrees of freedom due to physical constraints. The Stokes vector can also be considered as an incoherent superposition of a completely polarised part and an unpolarised part[3]. Those parameters have been adopted in different polarimetric applications[16,20,22,23,32,80,81]. The polarisation angle (PA) and intensity of the linear SOP also can be defined, with respect to dipole orientation applications[154-156] (see Fig. 5a (iv)). For a beam generated via an incoherent light source (such as a LED), the Stokes vectors can be directly added by scalar calculation. Therefore, partially polarised light can be divided into two parts – fully polarised/depolarised components[3], i.e., $\mathbf{S}_{total} = \mathbf{S}_u + \mathbf{S}_p$; where $\mathbf{S}_u$ and $\mathbf{S}_p$ represent fully depolarised and polarised components respectively.

For biomedical and clinical applications, characterising the vectorial properties of the outgoing light with a fixed incident SOP also showed great potential for structure identification[22-24,32]. **Wu & Walsh** reported that Stokes vector analysis with circular polarised illumination can reveal structural information about tissue[157]. **Macdonald & Meglinski** showed that turbid tissue can be quantitatively analysed via Stokes vector measurement with an optical clearing technique[158]. **Qi** et al. proposed a method of Stokes vector analysis for *ex vivo* porcine tongue, stomach, kidney and other tissues based on circular polarised illumination[159]; the most useful information was provided by circular-depolarisation and linear-retardance, which can normally be provided via MM decomposition[52,58] (as Stokes vector projections shown in Fig. 5a (ii)). **Kunnen** et al. employed Stokes vector detection with circular and elliptical incident SOPs for differentiation between healthy and cancerous lung tissues specifically using a Poincaré sphere illustration[160] (as Stokes vector locations shown in Fig. 5a (iii)). Note that the circular SOP illumination is especially useful for biomedical analysis, as its effects are independent of the orientation of the anisotropic components that widely exist in biomedical specimens[20-24,32]. What is more, its strong polarisation memory effect with respect to tissue-induced Mie-scattering has also gained attention[161] (here the memory effect[162] means that circular polarisation can survive many more scattering events than linear polarisation due to excessive forward scattering, hence it has higher probability to maintain the original information when passing through turbid tissue consisting of Mie scattering particles that are comparable in size to the wavelength).

*Information extraction from the vector properties of the object*

Measurement of the full vector properties of biomedical targets requires illumination with multiple SOPs in combination with multiple analyzing SOPs[20-24]. As we have mentioned above, the individual MM elements lack clear physical meanings, or explicit associations with microstructures[20-24]. That is to say, vectorial characteristics of the object, like diattenuation, retardance, and depolarisation are encoded within the MM elements. For a complex optical system (like tissue), each MM element is always associated with more than one polarisation property. Hence, numerous MM decomposition methods were proposed to quantitatively characterise the optical and structural properties of the object[52-59]. One prevalent method is the Mueller matrix polar-decomposition (MMPD) proposed by **Lu & Chipman**[52], which has been used and validated in lots of applications for characterisation of biomedical or material samples[58,163-168] (see Fig. 5b (i)); **He** et al. put forward the Mueller matrix transformation (MMT) with validations using phantom experiments and Monte Carlo simulations[57]. Based on the MMT concept, more rotation invariant parameters were extracted from the MM and applied to biomedical sample characterisation[168-170] (see Fig. 5b (ii)); **Arteaga** and colleagues derived Mueller matrix anisotropy coefficients (MMAC) to describe the level of different kinds of anisotropy for different polarisation systems[56]. Furthermore, other decomposition schemes were also developed, such as MM differential decomposition[55], symmetric decomposition[60,61] and Cloude decomposition[62]. Among the various decomposition approaches, different mathematical assumptions need to be made for different applications[52-62], such as assuming a determined layer sequence of different fundamental polarisation components for a complex object, which in effect simplifies the matrix reciprocity problem[52]. Recently, those methods and related parameters have also been compared quantitatively with each other for the purpose of structural characterisation[171-173]. We can summarise the parameters derived via the above methods: MMPD: diattenuation (D), retardance (R) and depolarisation (Δ) (all of them maintain linear/circular components)[52]; MMT: depolarisation (1-b), level of linear anisotropy ($t_1$), diattenuation property ($t_2$), level of birefringence ($t_3$) and fast axis orientation ($x_3$) and more[16,57]; MMAC: horizontal linear anisotropy ($\alpha$), 45° linear anisotropy ($\beta$) and circular anisotropy ($\gamma$) respectively with respect to the global anisotropy of the MM[56].

The MM contains fundamental physical characters like polarisance, diattenuation, retardance and depolarisation (shown in Fig. 2c); however, some concepts like anisotropy can be a combination of several fundamental polarisation processes[16,57]. It is worth mentioning that the depolarisation property – which is used for evaluating a SOP's disorder, randomness, or uncertainty[3-5] – is also linked with the concept of entropy in polarimetric research[11,174]. While the above parameters are derived from a full MM (4 × 4); **Ghosh** et al. and **Wang** et al. also reported works on 3 × 3 MM decomposition methods, related simulations and experiments, with an emphasis on biomedical applications[175,176]. In summary, the decomposed linear depolarisation and linear retardance from a 3 × 3 MM display similar qualitative relationships to the changes with respect to the microstructure of the sample, such as the density, molecule size, and orientation distributions of the scatterers as well as birefringence level of the interstitial medium[175,176].

The MM decomposition methods all require different assumptions (strong or weak) such as matrix reciprocity, the order that polarisation effects happen in the media, or homogeneity for the tissue analysis[52,58,177-179]. Therefore, their decomposed values are not strictly physically determined, if the assumptions do not hold in reality, which may well be the case, as biological tissue has high spatial complexity[58]. However, extraction through the MM polarisation parameters that have less assumptions and clearer physical meaning is always something to strive for. Several works pointed in such a direction: 1) **Gil** et al. and **Li** et al. proposed different polarisation parameters with physical determination via the asymmetric properties of the MM elements[169,170], by considering assumptions about layer constructions or the presence of absence of specific vector properties such as polarisance or diattenuation; 2) **Dong** et al. employed a data-driven machine learning technique to fit several polarimetry feature parameters (PFPs) for characterising determined pathological applications, such as detection of the abnormal areas of breast carcinoma and cervical cancerous tissue slices[95] (see Fig. 5b (iii)); 3) Breaking or restoring the symmetry (see Fig. 5b (iv)), based on analysis of different sub-regions of the MM, to extract determined information of the system is recently gaining interest[180]; The information extraction process is gradually developing from an analytical mathematics approach (equation-based, forward problem), to fitting or observing vectorial semantics/metrics (data-based, or shape/form-based inverse problem).

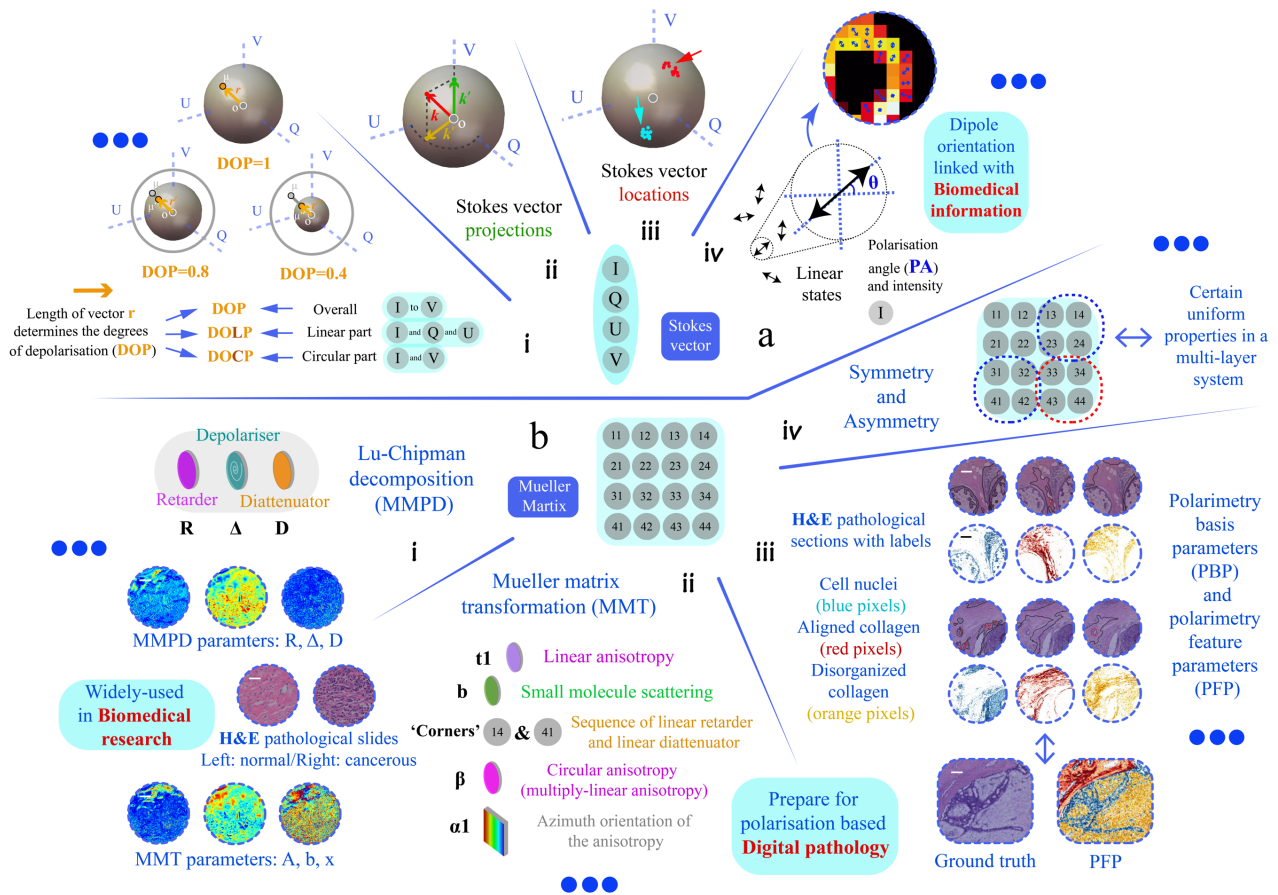

**Figure 5 Vectorial information extraction approaches.** (a) Stokes vector based approaches: (i) DOP, DOLP and DOCP[2-5]; (ii) Stokes vector projection approach[159]; (iii) Stokes vector location approach[160]; (iv) Dipole orientation differentiation approach. Polarisation angle (PA) and intensity are used as main parameters[155,156]. (b) MM based approaches: (i) MMPD method[52]: diattenuation (D), retardance (R) and depolarisation (Δ) (all of them maintain linear/circular components); and (ii) MMT method[57]: depolarisation (1-b; associated with small molecule scattering), level of linear anisotropy (t1), azimuth orientation of the anisotropy (α1), and more[180]; (iii) PFP method[95]; (iv) Property of symmetry and asymmetry[169,170]. (b(i)) Adapted with permission from Ref[168] © The Optical Society. (b(iii)) © [2021] IEEE. Reprinted, with permission, from Ref[95].

## 4 Vectorial information analysis for biomedical applications

Polarimetric techniques maintain unique advantages compared with other optical techniques: they can provide extra vectorial information through methods that are compatible with many existing optical systems, such as microscopes and endoscopes[16,24,32,33,92,181]. Much existing biomedical polarimetry research concerns sensing of bio-information in a label-free way without extraneous dyes[16,22,24]. In other areas, polarimetry can be used to characterise the vectorial information of fluorescence dyes, as the dipole orientation of the fluorophore is encoded in the polarisation state of the emitted light[155,156]. The SOP of such emission is always in a linear state; hence the polarisation angle (PA) and intensity of the linear SOP are quantities that can be harnessed, such as in biomedical applications in super-resolution microscopy[154,182,183]. Here we briefly summarize common phantoms used for biomedical polarimetric techniques. These techniques include: polarised wide-field microscopy[16,24,184], polarised light spatial frequency imaging[185], polarimetric endoscopy[186-191], spectral light scattering polarimetry[18,82,192-194], polarised fluorescence spectroscopy[195-197], polarised confocal microscopy[198], polarised Raman-spectroscopy[199,200], polarised super-resolution microscopy[155,156], polarisation sensitive optical coherence tomography[201-219], non-diffraction beam polarimetry (such as Bessel beam based)[220], polarisation-resolved nonlinear microscopy (including second/third harmonic generation)[221-227], and polarised speckle imaging[214,228] (several techniques will be mentioned again in the Discussion). The relationship between incoherence and depolarisation of the light should be kept in mind when considering coherence based polarimetric techniques: they are different but related optical concepts. If a polarised coherent beam passing through a scattering medium becomes incoherent, it can result in either polarised light or depolarised light. If after such a medium a polarised coherent beam changes into depolarised, the coherence property may still be maintained. For more details see Ref[229]. Several of the above techniques have also been adopted in three-dimensional (3D) imaging with signal integrations or sample segmentations[230]. However, numerous existing polarimetry techniques (within the scope of this review) fall into two-dimensional (2D) analysis[23-29]. With the completion of the cutting-edge mathematical interpretations and methodologies (see Discussion) there exists of course intriguing scope for further explorations.

In order to understand the interactions between polarised photons and biological specimens, and link the parameters obtained via the Stokes vector or MM with the biomedical microstructural information, a software phantom – Monte Carlo (MC) simulation – was proposed to give plausible explanations for the originality of the observed physical phenomena[45,46]. While biomedical samples are considered as turbid media with complex structures, different fundamental units to mimic the microstructural architecture have been employed: spherical scatterers[35,46]; cylindrical scatterers[41,46]; birefringent intermedia[37-39], multi-layered geometry[45] and so on[46]. MC simulations have successfully reproduced most of the important polarimetric characteristic features for biomedical samples[16,231,232].

*Thin specimens*

Specimens and their mimicking phantoms can be thin or bulky, which also in general determines the configurations of the biomedical polarimetry. A transmissive geometry is used for the thin cases (see Fig. 6) which are less scattering, thus most of the incident photons would be transmitted. A backscattering geometry (see Fig. 7) is preferred for the bulk cases (*ex vivo* and *in vivo*) which are highly scattering and depolarising, thus most of the incident photons would be backscattered. There is no clear boundary between what constitutes thin or bulk tissues. Indeed, intermediate or mixed states can exist, for which both the transmission and backscattering photons can be detected simultaneously[16,22,23,153]. In response to the beginning of this section, biomedical polarimetry can be used in labelled or label-free measurement; Figure 6 gives a summary for two types of the use of thin tissue polarimetry.

For label-based direction, polarimetry has found use in scientific applications, such as biomedical microscopy[16]. The vectorial information of the dipole emitters is encoded in the SOP of the detected light[154,155]. The dipole orientation (and the fluorescence intensity) polarimetric detection technique plays an important role in thin biomedical sample analysis: e.g., in fluorescence polarisation microscopy (FPM)[195,196,233,234]; FPM can be used to study the nuclear pore complex subcomplexes and the relative orientations[235], or be used to study different types of cytoskeleton such as actin, myosin, kinesin, microtubule and septin – those closely related with the performance of the dipole behaviours[236-240] – enabling research such as ATP and ADP binding[238]. Advanced research has been adopted in super-resolution imaging harnessing fluorescent dipoles via polarised illumination, with applications such as revealing heterogeneity and dynamics of subcellular lipid membranes[182,241,242]. These fluorescence anisotropy properties also belong to the fundamental polarisation properties that are encoded in the MM.

For label-free biomedical polarimetric research, especially in clinical/pathological related topics, cancerous tissues detection is an important application[22-24]. In the past decades, such polarimetric techniques have assisted the diagnosis of various cancerous tissues, such as human skin cancer[243], cervical cancer[244-247], colon cancer[167,248-251], liver cancer[164,252], breast cancer and gastrointestinal cancer[93-95,253]. A typical bio-information analysis of polarimetric data is for quantitative evaluation of the fibrosis process among different stages of cancer development[94,164]. Beside the degree of fibrosis that can be quantified via biomedical polarimetry, the distribution of features in the fibrous regions also can serve as another characteristic parameter to assist the pathological diagnosis; this distribution can be readily extracted via polarisation information[165,166,172]. Intuitively, such structures contribute intrinsic birefringence mainly affecting the fourth row and fourth column of the target MM[16]. A good demonstration in Ref[165] shows how polarimetric textural mapping of retardance properties can distinguish between the Crohn's disease and gastrointestinal luminal tuberculosis tissues (see Fig. 6b). Some thin specimen phantoms, as found in Ref[153], target the fundamental understanding of the constitution of certain biomedical specimens, such as using nano-particles or microspheres. Moreover, polarimetry has recently been applied to other diseases detections including Alzheimer's disease and bladder outlet obstruction[24,254,255].

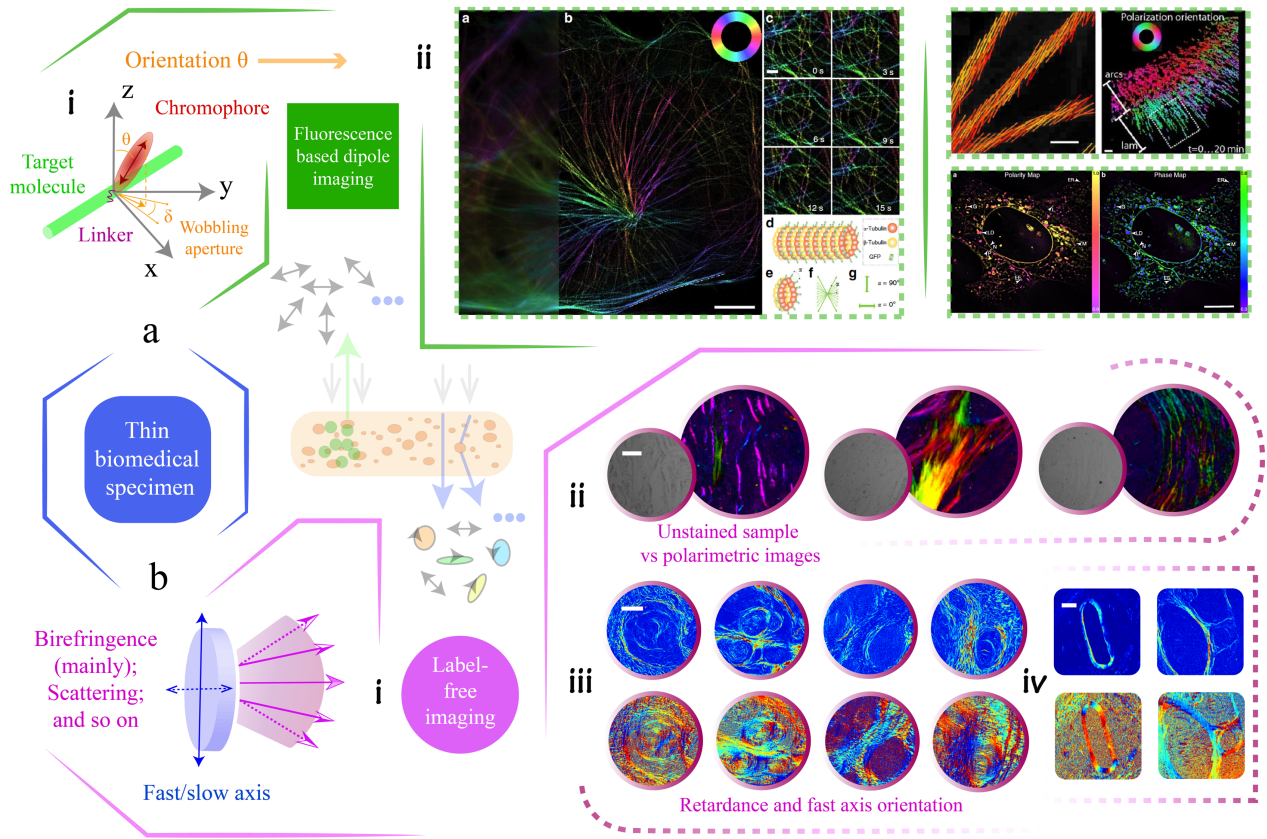

**Figure 6 Two categories of the applications enabled via thin specimen polarimetry.** (a) (i) Model of the dipole and target molecule with linker – related fundamental mechanism can be found in Ref[155,156]. (ii) Fluorescence dipole orientation imaging[182,183,234,256]. The polarisation orientation is given as a demonstration. (b) (i) Model of the main properties in thin specimen for label-free polarimetric imaging. (ii) Unstained/polarimetric imaging of human cervical and liver carcinoma tissue samples[32]. (iii) Polarimetric parameters imaging results to distinguish between Crohn's disease and gastrointestinal luminal tuberculosis tissues[165]. (iv) Polarimetric parameters images of human liver cirrhosis samples in different stages[164]. (a(ii) besides upper-right) Adapted with permission from Ref[182,183]. CC BY 4.0. (a(ii); upper-right) Adapted from Ref[234,256]. (b(ii) and b(iv)) Adapted from Ref[32,164]. CC BY 4.0. (b(iii)) Copyright Wiley-VCH GmbH. Reproduced with permission Ref[165].

***Bulk specimens***

Polarisation techniques can help improve the image contrast of the superficial layers of tissues by eliminating multiply scattered photons from the deep layers[20-24]. The previous literature shows that more than 85% of cancers originate from the superficial epithelium, which means that polarisation imaging methods have great potential in screening and identifying cancer at an early stage[257]. This would be specifically useful for *in vivo* clinical diagnosis, such as for minimally invasive surgery (MIS)[24]. Measurements in *ex vivo* thin tissue can use a transmissive geometry, whereas *ex/in vivo* bulk tissue detection would need backscattering configurations. Figure 7 gives a brief demonstration of certain current research topics related to bulk tissue polarimetry.

For polarimetric bulk tissue research, *ex vivo* detection plays an important role[22-24]. For example, collagen fibres, which widely exist in tissues and organs such as tendons, skin and bladder (from porcine, swine, lobster, calf or other animals[157,258-261]), skeletal and myocardial muscle fibres[262-264], and elastin fibres are widely-used due to their linear birefringence properties[20-23,165]. The alignment directions of all such fibrous structures are also linked with the fast axis orientation of the generated linear birefringence[16,166,172]. Furthermore, the scattering of bulk media is also studied via the extracted depolarisation[16,168,265]. The retardance and depolarisation related properties are the dominant parts of the vectorial properties of bulk tissues, as the magnitude of diattenuation for majority of tissue is typically very small[159], with several exceptions like skeletal and myocardial muscles[168] (see Fig. 7a (i)). Previous research analysing muscle tissue[266] showed lower retardance compared with tendon tissue, owing to the cellularity of these tissues. Sections of the bulk myocardial fibre tissues showed two circularly aligned ring-shaped fibrous structures (see Fig. 7a (i)), revealing their anisotropic properties[168]. The different anisotropic vectorial information obtained from polarimetric measurements can be very helpful for the discrimination and identification of different fibrous structures in tissues[165,166,168].

While *ex vivo* studies are mainly oriented towards fundamental research[24,153,172] (e.g., understanding the vectorial properties characterisation; see Fig. 7a (i), 7a (ii), 7b (i)), *in vivo* bulk tissue polarimetry is geared towards applications[24,258,267]. Typical backscattering mode polarimetry includes polarisation endoscopy[24], reflection MM microscopy[16,268], MM colposcopy[181,269], wide-field handheld polarimetry[16], and PS-OCT[270] (see Fig. 7a (iii), 7b (ii)), targeted to clinical diagnosis *in vivo*. As a promising *in vivo*, label-free diagnostic tool, polarisation endoscopic imaging has been implemented inside rat abdomen, revealing the small bowel, stomach, liver and fat with different polarisation characters[92]. Recent work also includes development of several different types of MM endoscope[189-191,258], and extension into the spectral domain (with certain fixed wavelengths)[92] (see Fig. 7a (iii)). PS-OCT[201-219] is specifically used for *in vivo* ophthalmic imaging, where polarimetric data accompanied with clinical analysis has been demonstrated, for retinal imaging[270] (see Fig. 7b (ii)). Other types of bulk tissue analysis, such as human lung cancerous tissue[160] and skin tissue[271], show good prospects for future clinical diagnosis[16,22-24].

While thin samples can feature multiple scattering process, such processes are of course more significant in bulk samples[20-22]. Considering the intriguing scope of the polarimetric technique for *in vivo* clinical diagnosis, beside the simulations, there is a need for complex phantoms, such as those exhibiting birefringence (see Fig. 7b (iii)) or depolarisation, to establish reliable processes for investigation of complex scattering mechanisms[272,273]. A recent review has summarised various phantoms for both thin and bulk samples[153]. Micro-spheres, silicon-based phantoms, nanoparticles, cylindrical scatterers, and birefringent/dichroism films[35,159,272,274] have all been employed in various validations. *In vivo* biomedical polarimetry and its related applications clearly offer a large space for future exploration.

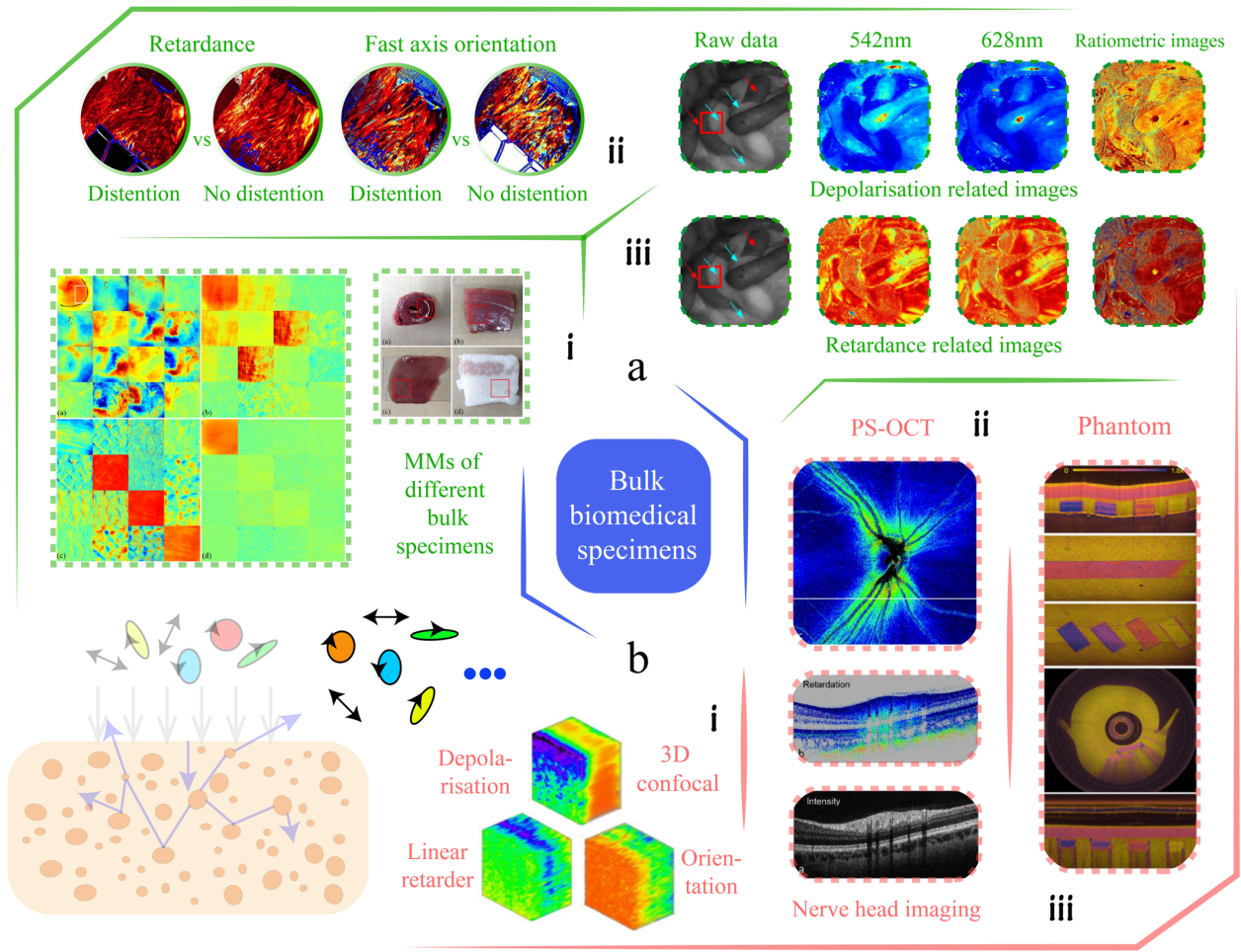

**Figure 7 Applications enabled via bulk specimen polarimetry (*in vivo* and *ex vivo*).** (a) General backscattering mode polarimetric imaging: (i) MMs and original sample images for several ex vivo bulk specimens[168]. (ii) Polarisation properties of *ex vivo* bladder tissue[258]. (iii) Polarisation properties using combined wavelength information for *in vivo* characterisation of rat abdomen tissue[92]. (b) Certain specific backscattering-mode phantoms: (i) Full-depth MM confocal imaging of an unstained rat cornea[267]. (ii) PS-OCT imaging for a nerve head, which is conducted under the condition of *in vivo* eye imaging[34]. (iii) A birefringent phantom designed for bulk tissue research[272]. (a(ii)) Adapted with permission Ref[258]. CC BY 4.0. (a(i), a(iii), b(i), b(ii) and b (iii)) Adapted with permission from Ref[34,92,168,267,272] © The Optical Society.

## 5 Directions for advanced biomedical polarimetry and future prospects

Biomedical applications of polarimetry have attracted substantial attention. We hope this short review paper gives readers a general overview from fundamental polarisation concepts, through polarimetric techniques, to recent biomedical and clinical applications[7,16,20-24,29,34,63,79,153]. In addition to the summaries of recent research trends explained above, we provide here some further perspective on prospects in this application area, considering the use of polarimetry in a multimodal combination with other advanced technologies (see Fig. 8 for a summary).

Firstly, the fast development of the machine learning (ML) is clearly going to have an impact on this field[95,275,276]. Such data-driven techniques may pave new directions for biomedical polarimetry, either through improving the quality of polarimetry (such as overcoming the numerous sources of error) or through enhanced information extraction[68,95]. One possibility is to use low resolution information to reconstruct high resolution patterns (following the spirit of works such as Ref[277,278]). Secondly, while ML is geared towards improving the information processing aspects of polarimetry, new adaptive optics techniques can be used to extend the capabilities of polarimetry through full vectorial beam control. This could enable enhanced polarisation imaging resolution physically via beam shaping and compensation of polarisation errors[279-282]. Thirdly, the emerging techniques based on metasurfaces – subwavelength arrays of nano-scatterers that can modify polarisation – have been adopted for polarimetry[283], as well as for 3D polarisation control[284]. Such developments may bring new opportunities for advanced biomedical polarimetry, such as forming compact vectorial sensors[24,285] for deep tissue information extraction. Fourthly, second harmonic generation (SHG) and third harmonic generation (THG) based 3D MM techniques have been proposed[286-289]. These are described by extended MMs that are more complicated than $4 \times 4$ MMs used for linear scattering ($4 \times 9$ and $4 \times 16$ elements, respectively, for SHG and THG)[286,287]. For these methods, further advanced information extraction and analysis approaches are of course intriguing. Finally, the intensity and wavelength have been utilized together with polarisation in polarimetry for a long time. However, the absolute phase information – especially geometric phase-related techniques[8,10] – may again open windows for new biomedical polarimetry approaches with multi-modal performance.

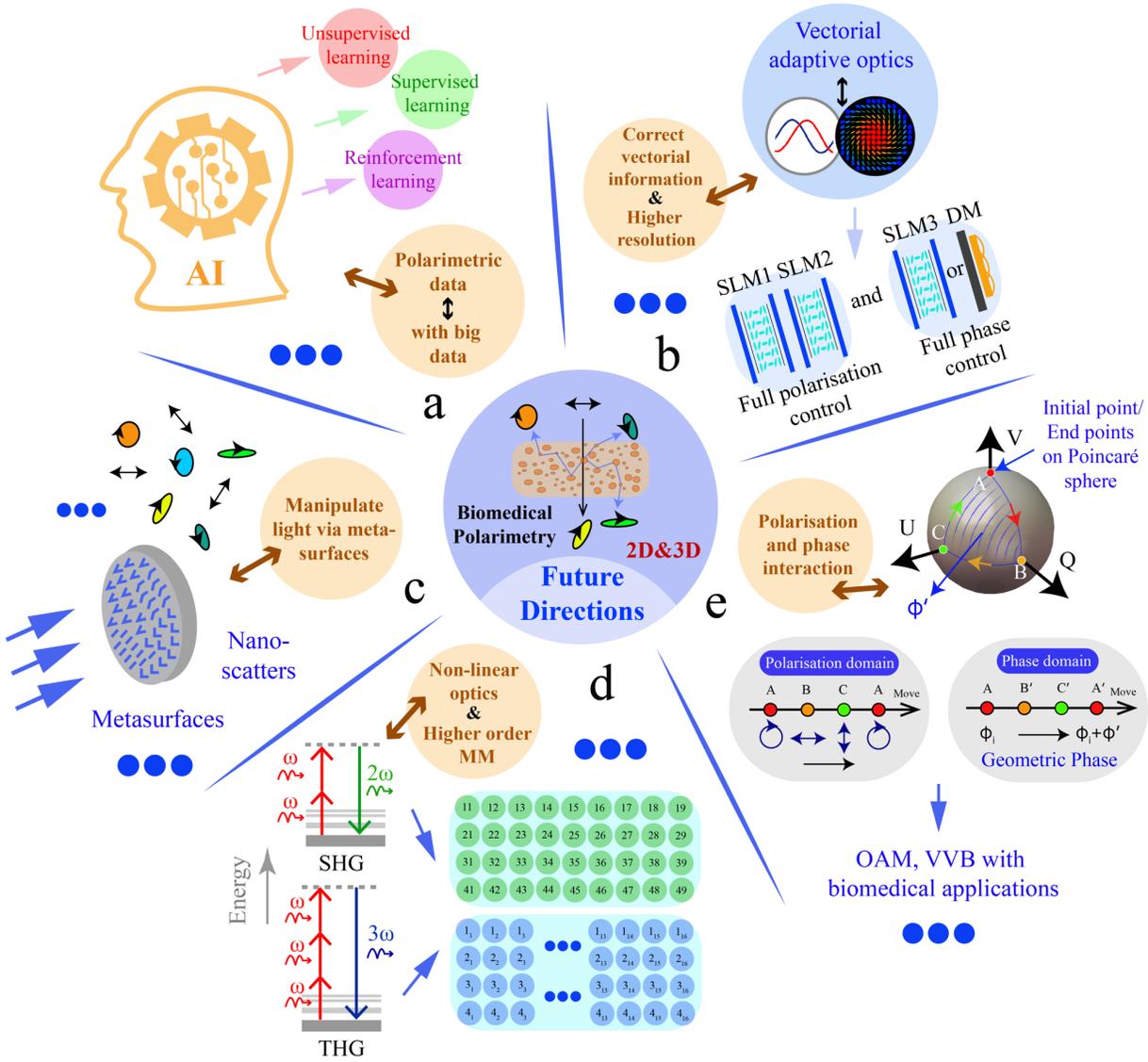

**Figure 8 Future directions for biomedical polarimetry.** (a) Combination with machine learning techniques and big data. (b) Combination with vectorial adaptive optics. (c) Combination with metasurface based techniques. (d) Combination with non-linear techniques, such as SHG and THG, in which a high order MM is required. (e) Combination with absolute phase information, such as geometric phase, with potential applications related with orbital angular momentum (OAM) and vector vortex beam (VVB) manipulations.

**Acknowledgments**

The project was supported by the European Research Council (AdOMiS, no. 695140); Shenzhen Fundamental Research and Discipline Layout Project (JCYJ20170412170814624); The first author C.H. (also the figure maker) would like to thank Prof. Jessica C Ramella-Roman at Florida International University, Prof. Daniel S Elson at Imperial College London, Prof. Christoph K Hitzenberger at Medical University of Vienna, Prof. Johannes Fitzgerald de Boer at Vrije Universiteit Amsterdam, Prof. Yoshiaki Yasuno at University of Tsukuba, Prof. Martin Villiger at Harvard Medical School, Prof. Nan Zeng and Prof. Yonghong He at Tsinghua University, Dr. Long Chen at University of Oxford for useful discussions and/or support through provision of the original graphics of their works.

**Competing interests**

The authors declare no competing interests.

**Additional Information**

Correspondence and request for materials should be addressed to C.H., H.H., or M.J.B.

**References**

1  Ronchi, V. & Barocas, V. The nature of light: An historical survey. *The nature of light: An historical survey* (1970).

2  Huard, S. *Polarisation of light* (1997).

3  Goldstein, D. H. *Polarised light* (CRC press, 2017).

4  Chipman, R. A., Lam, W. S. T. & Young, G. *Polarised light and optical systems* (CRC press, 2018).

5  Pérez, J. J. G. & Ossikovski, R. *Polarised light and the Mueller matrix approach* (CRC press, 2017).

6 Zhan, Q. W. Cylindrical vector beams: from mathematical concepts to applications. *Advances in Optics and Photonics* **1**, 1-57 (2009).

7 Rosales-Guzmán, C., Ndagano, B. & Forbes, A. A review of complex vector light fields and their applications. *Journal of Optics* **20**, 123001 (2018).

8 Forbes, A., de Oliveira, M. & Dennis, M. R. Structured light. *Nature Photonics* **15**, 253-262 (2021).

9 Wang, J., Castellucci, F. & Franke-Arnold, S. Vectorial light–matter interaction: Exploring spatially structured complex light fields. *AVS Quantum Science* **2**, 031702 (2020).

10 Slussarenko, S. *et al.* Guiding light via geometric phases. *Nature Photonics* **10**, 571-575 (2016).

11 Cloude, S. *Polarisation: applications in remote sensing* (OUP Oxford, 2009).

12 Ndagano, B. *et al.* Characterising quantum channels with non-separable states of classical light. *Nature Physics* **13**, 397-402 (2017).

13 Marrucci, L., Manzo, C. & Paparo, D. Optical spin-to-orbital angular momentum conversion in inhomogeneous anisotropic media. *Physical Review Letters* **96**, 163905 (2006).

14 Bliokh, K. Y., Rodriguez-Fortuno, F. J., Nori, F. & Zayats, A. V. Spin-orbit interactions of light. *Nature Photonics* **9**, 796-808 (2015).

15 Schulz, M. *et al.* Giant intrinsic circular dichroism of prolinol-derived squaraine thin films. *Nature Communications* **9**, 2413 (2018).

16 He, H. *et al.* Mueller matrix polarimetry—an emerging new tool for characterising the microstructural feature of complex biological specimen. *Journal of Lightwave Technology* **37**, 2534-2548 (2018).

17 Oldenbourg, R. A new view on polarisation microscopy. *Nature* **381**, 811-812 (1996).

18 Gurjar, R. S. *et al.* Imaging human epithelial properties with polarised light-scattering spectroscopy. *Nature Medicine* **7**, 1245-1248 (2001).

19 Qiu, L. *et al.* Multispectral scanning during endoscopy guides biopsy of dysplasia in Barrett's esophagus. *Nature Medicine* **16**, 603-606 (2010).

20 Ghosh, N. & Vitkin, I. A. Tissue polarimetry: concepts, challenges, applications, and outlook. *Journal of Biomedical Optics* **16**, 110801 (2011).

21	Novikova, T., Meglinski, I., Ramella-Roman, J. C. & Tuchin, V. V. Special Section Guest Editorial: Polarised Light for Biomedical Applications. *Journal of Biomedical Optics* **21**, 71001 (2016).

22	Tuchin, V. V. Polarised light interaction with tissues. *Journal of Biomedical Optics* **21**, 71114 (2016).

23	Ramella-Roman, J., Saytashev, I. & Piccini, M. A review of polarisation-based imaging technologies for clinical and pre-clinical applications. *Journal of Optics* **22**, 123001 (2020).

24	Qi, J. & Elson, D. S. Mueller polarimetric imaging for surgical and diagnostic applications: a review. *Journal of Biophotonics* **10**, 950-982 (2017).

25	Fujiwara, H. *Spectroscopic ellipsometry: principles and applications* (John Wiley & Sons, 2007).

26	Azzam, R. M., Bashara, N. M. & Ballard, S. S. Ellipsometry and polarised light. *Physics Today* **31**, 72 (1978).

27	Mendoza-Galván, A. *et al.* Mueller matrix spectroscopic ellipsometry study of chiral nanocrystalline cellulose films. *Journal of Optics* **20**, 024001 (2018).

28	Azzam, R. M. Stokes-vector and Mueller-matrix polarimetry. *Journal of Optical Society of America A* **33**, 1396-1408 (2016).

29	Azzam, R. M. in *Polarisation: Measurement, Analysis, and Remote Sensing* 396-405 (International Society for Optics and Photonics).

30	Song, B. K. *et al.* Broadband optical properties of graphene and HOPG investigated by spectroscopic Mueller matrix ellipsometry. *Applied Surface Science* **439**, 1079-1087 (2018).

31	Jiang, H. *et al.* Characterisation of volume gratings based on distributed dielectric constant model using Mueller matrix ellipsometry. *Applied Sciences-Basel* **9**, 698 (2019).

32	Chang, J. T. *et al.* Division of focal plane polarimeter-based 3×4 Mueller matrix microscope: a potential tool for quick diagnosis of human carcinoma tissues. *Journal of Biomedical Optics* **21**, 056002 (2016).

33	Fu, Y. *et al*. Flexible 3×3 Mueller matrix endoscope prototype for cancer detection. *IEEE Transactions on Instrumentation and Measurement* **67**, 1700-1712 (2018).

34	de Boer, J. F., Hitzenberger, C. K. & Yasuno, Y. Polarisation sensitive optical coherence tomography - a review. *Biomedical Optics Express* **8**, 1838-1873 (2017).

35	Ghosh, N., Patel, H. S. & Gupta, P. K. Depolarisation of light in tissue phantoms - effect of a distribution in the size of scatterers. *Optics Express* **11**, 2198-2205 (2003).

36   Kienle, A. & Hibst, R. Light guiding in biological tissue due to scattering. *Physical Review Letters* **97**, 018104 (2006).

37   Baravian, C., Dillet, J. & Decruppe, J. P. Birefringence determination in turbid media. *Physical Review E* **75**, 032501 (2007).

38   Wang, X. & Wang, L. V. Propagation of polarised light in birefringent turbid media: a Monte Carlo study. *Journal of Biomedical Optics* **7**, 279-290 (2002).

39   Wang, X. & Wang, L. Propagation of polarised light in birefringent turbid media: time-resolved simulations. *Optics Express* **9**, 254-259 (2001).

40   Du, E. *et al.* Two-dimensional backscattering Mueller matrix of sphere-cylinder birefringence media. *Journal of Biomedical Optics* **17**, 126016, (2012).

41   He, H. *et al.* Application of sphere-cylinder scattering model to skeletal muscle. *Optics Express* **18**, 15104-15112 (2010).

42   Chen, D. *et al.* Mueller matrix polarimetry for characterising microstructural variation of nude mouse skin during tissue optical clearing. *Biomedical Optics Express* **8**, 3559-3570, (2017).

43   Donner, C. & Jensen, H. W. Light diffusion in multi-layered translucent materials. *Acm Transactions on Graphics* **24**, 1032-1039 (2005).

44   Hulst, H. C. & van de Hulst, H. C. *Light scattering by small particles* (Courier Corporation, 1981).

45   Wang, L., Jacques, S. L. & Zheng, L. MCML—Monte Carlo modeling of light transport in multi-layered tissues. *Computer methods and programs in biomedicine* **47**, 131-146 (1995).

46   Yun, T. *et al.* Monte Carlo simulation of polarised photon scattering in anisotropic media. *Optics Express* **17**, 16590-16602 (2009).

47   Brosseau, C. *Fundamentals of polarised light: a statistical optics approach* (Wiley-Interscience, 1998).

48   Born, M. & Wolf, E. *Principles of optics: electromagnetic theory of propagation, interference and diffraction of light* (Elsevier, 2013).

49   Ling, X. *et al.* Recent advances in the spin Hall effect of light. *Rep Prog Phys* **80**, 066401 (2017).

50   Rubinsztein-Dunlop, H. *et al.* Roadmap on structured light. *Journal of Optics* **19**, 013001 (2016).

51  Bass, M. *et al.* Handbook of Optics, Volume IV: Optical Properties of Materials. *Nonlinear Optics, Quantum Optics* **4** (2009).

52  Lu, S. Y. & Chipman, R. A. Interpretation of Mueller matrices based on polar decomposition. *Journal of the Optical Society of America A* **13**, 1106-1113 (1996).

53  Arteaga, O. & Canillas, A. Pseudopolar decomposition of the Jones and Mueller-Jones exponential polarisation matrices. *Journal of the Optical Society of America A* **26**, 783-793 (2009).

54  Ossikovski, R., De Martino, A. & Guyot, S. Forward and reverse product decompositions of depolarising Mueller matrices. *Optics Letters* **32**, 689-691 (2007).

55  Ortega-Quijano, N. & Arce-Diego, J. L. Mueller matrix differential decomposition. *Optics Letters* **36**, 1942-1944 (2011).

56  Arteaga, O., Garcia-Caurel, E. & Ossikovski, R. Anisotropy coefficients of a Mueller matrix. *Journal of the Optical Society of America A* **28**, 548-553 (2011).

57  He, H. *et al.* A possible quantitative Mueller matrix transformation technique for anisotropic scattering media/Eine mögliche quantitative Müller-Matrix-Transformations-Technik für anisotrope streuende Medien. *Photonics & Lasers in Medicine* **2**, 129-137 (2013).

58  Ghosh, N., Wood, M. F. & Vitkin, I. A. Mueller matrix decomposition for extraction of individual polarisation parameters from complex turbid media exhibiting multiple scattering, optical activity, and linear birefringence. *Journal of Biomedical Optics* **13**, 044036 (2008).

59  Gil, J. J. Characteristic properties of Mueller matrices. *Journal of the Optical Society of America A* **17**, 328-334 (2000).

60  Ossikovski, R. Analysis of depolarising Mueller matrices through a symmetric decomposition. *Journal of the Optical Society of America A* **26**, 1109-1118 (2009).

61  Vizet, J. & Ossikovski, R. Symmetric decomposition of experimental depolarising Mueller matrices in the degenerate case. *Applied Optics* **57**, 1159-1167 (2018).

62  Cloude, S. R. Group-theory and polarisation algebra. *Optik* **75**, 26-36 (1986).

63  Pezzaniti, J. L. & Chipman, R. A. Mueller matrix imaging polarimetry. *Optical Engineering* **34**, 1558-1568 (1995).

64  Azzam, R. M. Photopolarimetric measurement of the Mueller matrix by Fourier analysis of a single detected signal. *Optics Letters* **2**, 148 (1978).

65  Goldstein, D. H. Mueller matrix dual-rotating retarder polarimeter. *Applied Optics* **31**, 6676-6683 (1992).

66  Smith, M. H. Optimization of a dual-rotating-retarder Mueller matrix polarimeter. *Applied Optics* **41**, 2488-2493 (2002).

67  Dubreuil, M., Rivet, S., Le Jeune, B. & Cariou, J. Snapshot Mueller matrix polarimeter by wavelength polarisation coding. *Optics Express* **15**, 13660-13668 (2007).

68  He, C. *et al.* Full Poincare mapping for ultra-sensitive polarimetry. *arXiv preprint arXiv:2101.09372* (2021).

69  Sabatke, D. S. *et al.* Optimization of retardance for a complete Stokes polarimeter. *Optics Letters* **25**, 802-804 (2000).

70  Tyo, J. S. Design of optimal polarimeters: maximization of signal-to-noise ratio and minimization of systematic error. *Applied Optics* **41**, 619-630 (2002).

71  He, C. *et al.* Linear polarisation optimized Stokes polarimeter based on four-quadrant detector. *Applied Optics* **54**, 4458-4463 (2015).

72  He, C. *et al.* Complex vectorial optics through gradient index lens cascades. *Nature Communications* **10**, 1-8 (2019).

73  Li, X. *et al.* Learning-based denoising for polarimetric images. *Optics Express* **28**, 16309-16321 (2020).

74  Abubakar, A. *et al.* Hybrid denoising algorithm of BM3D and KSVD for Gaussian noise in DoFP polarisation images. *IEEE Access* **8**, 57451-57459 (2020).

75  Bueno, J. M. Polarimetry using liquid-crystal variable retarders: theory and calibration. *Journal of Optics A* **2**, 216-222 (2000).

76  Skumanich, A., Lites, B. W., Pillet, V. M. & Seagraves, P. The calibration of the advanced Stokes polarimeter. *Astrophysical Journal Supplement Series* **110**, 357-380 (1997).

77  Arteaga, O., Freudenthal, J., Wang, B. & Kahr, B. Mueller matrix polarimetry with four photoelastic modulators: theory and calibration. *Applied Optics* **51**, 6805-6817 (2012).

78  Smith, M. H. *et al.* in *Polarisation Analysis, Measurement, and Remote Sensing III.* 55-64 (International Society for Optics and Photonics).

79  Tyo, J. S., Goldstein, D. L., Chenault, D. B. & Shaw, J. A. Review of passive imaging polarimetry for remote sensing applications. *Applied Optics* **45**, 5453-5469 (2006).

80  Jacques, S. L., Ramella-Roman, J. C. & Lee, K. Imaging skin pathology with polarised light. *Journal of Biomedical Optics* **7**, 329-340 (2002).

81  Jacques, S. L., Roman, J. R. & Lee, K. Imaging superficial tissues with polarised light. *Lasers in Surgery and Medicine* **26**, 119-129 (2000).

82  Demos, S. G., Radousky, H. B. & Alfano, R. R. Deep subsurface imaging in tissues using spectral and polarisation filtering. *Optics Express* **7**, 23-28 (2000).

83  Demos, S. & Alfano, R. Optical polarisation imaging. *Applied optics* **36**, 150-155 (1997).

84  Groner, W. *et al.* Orthogonal polarisation spectral imaging: a new method for study of the microcirculation. *Nature Medicine* **5**, 1209-1212 (1999).

85  Bargo, P. R. & Kollias, N. Measurement of skin texture through polarisation imaging. *British Journal of Dermatology* **162**, 724-731 (2010).

86  Sridhar, S. & Da Silva, A. Enhanced contrast and depth resolution in polarisation imaging using elliptically polarised light. *Journal of Biomedical Optics* **21**, 071107 (2016).

87  Collett, E. Measurement of the four Stokes polarisation parameters with a single circular polariser. *Optics Communications* **52**, 77-80 (1984).

88  Laude-Boulesteix, B., De Martino, A., Drevillon, B. & Schwartz, L. Mueller polarimetric imaging system with liquid crystals. *Applied Optics* **43**, 2824-2832 (2004).

89  Sornsin, E. A. & Chipman, R. A. Polarimetric Mueller matrix polarimetry of electro-optic PLZT spatial light modulators. *Proceedings of SPIE* **2873**, 196-201 (1996).

90  Peinado, A., Lizana, A. & Campos, J. Optimization and tolerance analysis of a polarimeter with ferroelectric liquid crystals. *Applied Optics* **52**, 5748-5757 (2013).

91  Alali, S., Gribble, A. & Vitkin, I. A. Rapid wide-field Mueller matrix polarimetry imaging based on four photoelastic modulators with no moving parts. *Optics Letters* **41**, 1038-1041 (2016).

92  Qi, J. *et al.* Narrow band 3 × 3 Mueller polarimetric endoscopy. *Biomedical Optics Express* **4**, 2433-2449 (2013).

93    Dong, Y., Liu, S., Shen, Y., He, H. & Ma, H. Probing variations of fibrous structures during the development of breast ductal carcinoma tissues via Mueller matrix imaging. *Biomedical Optics Express* **11**, 4960-4975 (2020).

94    Dong, Y. *et al.* Quantitatively characterising the microstructural features of breast ductal carcinoma tissues in different progression stages by Mueller matrix microscope. *Biomedical Optics Express* **8**, 3643-3655 (2017).

95    Dong, Y. *et al.* Deriving polarimetry feature parameters to characterise microstructural features in histological sections of breast tissues. *IEEE Transactions on Biomedical Engineering* **68**, 881-892 (2021).

96    Hagen, N., Oka, K. & Dereniak, E. L. Snapshot Mueller matrix spectropolarimeter. *Optics Letters* **32**, 2100-2102 (2007).

97    Azzam, R. Division-of-amplitude photopolarimeter (DOAP) for the simultaneous measurement of all four Stokes parameters of light. *Optica Acta: International Journal of Optics* **29**, 685-689 (1982).

98    Jellison, G. E., Jr. Four-channel polarimeter for time-resolved ellipsometry. *Optics Letters* **12**, 766-768 (1987).

99    Compain, E. & Drevillon, B. Broadband division-of-amplitude polarimeter based on uncoated prisms. *Applied Optics* **37**, 5938-5944 (1998).

100   Peinado, A. *et al.* Conical refraction as a tool for polarisation metrology. *Optics Letters* **38**, 4100-4103 (2013).

101   Haigh, J. A., Kinebas, Y. & Ramsay, A. J. Inverse conoscopy: a method to measure polarisation using patterns generated by a single birefringent crystal. *Applied Optics* **53**, 184-188 (2014).

102   Chang, J., Zeng, N., He, H., He, Y. & Ma, H. Single-shot spatially modulated Stokes polarimeter based on a GRIN lens. *Optics Letters* **39**, 2656-2659 (2014).

103   Bhandari, P., Voss, K. J. & Logan, L. An instrument to measure the downwelling polarised radiance distribution in the ocean. *Opt Express* **19**, 17609-17620 (2011).

104   Pezzaniti, J. L. & Chenault, D. B. in *Polarisation Science and Remote Sensing II.* 58880V (International Society for Optics and Photonics).

105   Zimmerman, B. G., Ramkhalawon, R., Alonso, M. & Brown, T. G. in *Three-Dimensional and Multidimensional Microscopy: Image Acquisition and Processing XXI.* 894912 (International Society for Optics and Photonics).

106   Chun, C. S., Fleming, D. L. & Torok, E. in *Automatic Object Recognition IV.* 275-286 (International Society for Optics and Photonics).

107    Nordin, G. P., Meier, J. T., Deguzman, P. C. & Jones, M. W. Micropolariser array for infrared imaging polarimetry. *Journal of the Optical Society of America A* **16**, 1168-1174 (1999).

108    Andreou, A. G. & Kalayjian, Z. K. Polarisation imaging: principles and integrated polarimeters. *IEEE Sensors Journal* **2**, 566-576 (2002).

109    Chen, Z., Wang, X. & Liang, R. Calibration method of microgrid polarimeters with image interpolation. *Applied Optics* **54**, 995-1001 (2015).

110    Gao, S. & Gruev, V. Bilinear and bicubic interpolation methods for division of focal plane polarimeters. *Optics Express* **19**, 26161-26173 (2011).

111    Gao, S. & Gruev, V. Gradient-based interpolation method for division-of-focal-plane polarimeters. *Optics Express* **21**, 1137-1151 (2013).

112    Gruev, V., Perkins, R. & York, T. CCD polarisation imaging sensor with aluminum nanowire optical filters. *Optics Express* **18**, 19087-19094 (2010).

113    Hsu, W. L. *et al.* Polarisation microscope using a near infrared full-Stokes imaging polarimeter. *Optics Express* **23**, 4357-4368 (2015).

114    Liu, Y. *et al.* Complementary fluorescence-polarisation microscopy using division-of-focal-plane polarisation imaging sensor. *Journal of Biomedical Optics* **17**, 116001 (2012).

115    Millerd, J. *et al.* in *Fringe 2005* 640-647 (Springer, 2006).

116    Ratliff, B. M., LaCasse, C. F. & Tyo, J. S. Interpolation strategies for reducing IFOV artifacts in microgrid polarimeter imagery. *Optics Express* **17**, 9112-9125 (2009).

117    Ratliff, B. M. *et al.* Dead pixel replacement in LWIR microgrid polarimeters. *Optics Express* **15**, 7596-7609 (2007).

118    Tyo, J. S., LaCasse, C. F. & Ratliff, B. M. Total elimination of sampling errors in polarisation imagery obtained with integrated microgrid polarimeters. *Optics Letters* **34**, 3187-3189 (2009).

119    York, T. & Gruev, V. in *Infrared Technology and Applications XXXVII.* 80120H (International Society for Optics and Photonics).

120   York, T. *et al.* Bioinspired Polarisation Imaging Sensors: From Circuits and Optics to Signal Processing Algorithms and Biomedical Applications: Analysis at the focal plane emulates nature's method in sensors to image and diagnose with polarised light. *Proceedings of IEEE* **102**, 1450-1469 (2014).

121   Zhang, Z. *et al.* Nano-fabricated pixelated micropolariser array for visible imaging polarimetry. *Review of Scientific Instruments* **85**, 105002 (2014).

122   Zhao, X. *et al.* Patterned dual-layer achromatic micro-quarter-wave-retarder array for active polarisation imaging. *Optics Express* **22**, 8024-8034 (2014).

123   Oka, K. & Saito, N. in *Infrared Detectors and Focal Plane Arrays VIII.* 629508 (International Society for Optics and Photonics).

124   Suárez-Bermejo, J. C., de Sande, J. C. G., Santarsiero, M. & Piquero, G. Mueller matrix polarimetry using full Poincaré beams. *Optics and Lasers in Engineering* **122**, 134-141 (2019).

125   Azzam, R. M. A. Arrangement of four photodetectors for measuring the state of polarisation of light. *Optics Letters* **10**, 309-311 (1985).

126   Goldstein, D. H. & Chipman, R. A. Error analysis of a Mueller matrix polarimeter. *Journal of the Optical Society of America A* **7**, 693-700 (1990).

127   Ahmad, J. E. & Takakura, Y. Error analysis for rotating active Stokes-Mueller imaging polarimeters. *Optics Letters* **31**, 2858-2860 (2006).

128   Dai, H. & Yan, C. Measurement errors resulted from misalignment errors of the retarder in a rotating-retarder complete Stokes polarimeter. *Optics Express* **22**, 11869-11883 (2014).

129   Mu, T., Zhang, C., Li, Q. & Liang, R. Error analysis of single-snapshot full-Stokes division-of-aperture imaging polarimeters. *Optics Express* **23**, 10822-10835 (2015).

130   Macias-Romero, C. & Torok, P. Eigenvalue calibration methods for polarimetry. *Journal of the European Optical Society-Rapid Publications* **7**, 12004 (2012).

131   Compain, E., Poirier, S. & Drevillon, B. General and self-consistent method for the calibration of polarisation modulators, polarimeters, and Mueller-matrix ellipsometers. *Applied Optics* **38**, 3490-3502 (1999).

132   De Martino, A., Garcia-Caurel, E., Laude, B. & Drévillon, B. General methods for optimized design and calibration of Mueller polarimeters. *Thin Solid Films* **455**, 112-119 (2004).

133     Marenko, V. & Molebnaya, T. Optimization of stokes polarimeters employing a measurement of 4intensities. *Soviet Journal of optical technology* **57**, 452-455 (1990).

134     Ambirajan, A. & Look Jr, D. C. in *Polarisation Analysis and Measurement II.* 314-326 (International Society for Optics and Photonics).

135     Ambirajan, A. & Look Jr, D. C. Optimum angles for a polarimeter: part I. *Optical Engineering* **34**, 1651-1655 (1995).

136     Tyo, J. S. Optimum linear combination strategy for an N-channel polarisation-sensitive imaging or vision system. *Journal of the Optical Society of America A* **15**, 359-366 (1998).

137     Tyo, J. S. in *Polarisation Analysis, Measurement, and Remote Sensing III.* 65-74 (International Society for Optics and Photonics).

138     Tyo, J. S. Noise equalization in Stokes parameter images obtained by use of variable-retardance polarimeters. *Optics Letters* **25**, 1198-1200 (2000).

139     Azzam, R. M. A., Elminyawi, I. M., El-Saba, A, M. General analysis and optimization of the four-detector photopolarimeter. *Journal of the Optical Society of America A* **5**, 681 (1988).

140     Tyo, J. S. Considerations in polarimeter design. *Proceedings of SPIE* 4133 (2000).

141     Peinado, A. *et al.* Optimization and performance criteria of a Stokes polarimeter based on two variable retarders. *Optics Express* **18**(8): 9815 (2010).

142     Foreman, M. R., Favaro, A. & Aiello, A. Optimal frames for polarisation state reconstruction. *Physical Review Letters* **115**, 263901 (2015).

143     Foreman, M. R. & Goudail, F. On the equivalence of optimization metrics in Stokes polarimetry. *Optical Engineering* **58**, 082410 (2019).

144     Twietmeyer, K. M. & Chipman, R. A. Optimization of Mueller matrix polarimeters in the presence of error sources. *Optics Express* **16**: 11589-11603 (2008).

145     Beckley, A. M., Brown, T. G. & Alonso, M. A. Full Poincare beams. *Optics Express* **18**, 10777-10785 (2010).

146     Beckley, A. M. *Polarimetry and beam apodization using stress-engineered optical elements*. (University of Rochester, 2012).

147   Dewage, A. A. G. & Brown, T. in *Complex Light and Optical Forces XV.* 117010N (International Society for Optics and Photonics).

148   Vella, A. & Alonso, M. A. Optimal birefringence distributions for imaging polarimetry. *Optics Express* **27**, 36799-36814 (2019).

149   Vella, A. & Alonso, M. A. in *Progress in Optics* Vol. 65 231-311 (Elsevier, 2020).

150   Vella, A. J. *Description and applications of space-variant polarisation states and elements*. (University of Rochester, 2018).

151   Ramkhalawon, R. D., Brown, T. G. & Alonso, M. A. Imaging the polarisation of a light field. *Optics Express* **21**, 4106-4115 (2013).

152   Zimmerman, B. G. & Brown, T. G. Star test image-sampling polarimeter. *Optics Express* **24**, 23154-23161 (2016).

153   Chue-Sang, J. *et al.* Optical phantoms for biomedical polarimetry: a review. *Journal of Biomedical Optics* **24**, 1-12 (2019).

154   Zhanghao, K. *et al.* Super-resolution dipole orientation mapping via polarisation demodulation. *Light: Science & Applications* **5**, e16166 (2016).

155   Zhanghao, K. *et al*. Super-resolution fluorescence polarisation microscopy. *Journal of Innovative Optical Health Sciences* **11**, 1730002 (2018).

156   Chen, L. *et al.* Advances of super-resolution fluorescence polarisation microscopy and its applications in life sciences. *Computational and Structural Biotechnology Journal* **18**, 2209-2216 (2020).

157   Wu, P. J. & Walsh Jr, J. T. Stokes polarimetry imaging of rat‐tail tissue in a turbid medium using incident circularly polarised light. *Lasers in Surgery and Medicine* **37**, 396-406 (2005).

158   Macdonald, C. & Meglinski, I. Backscattering of circular polarised light from a disperse random medium influenced by optical clearing. *Laser Physics Letters* **8**, 324-328 (2011).

159   Qi, J. *et al.* Assessment of tissue polarimetric properties using Stokes polarimetric imaging with circularly polarised illumination. *Journal of Biophotonics* **11**, e201700139 (2018).

160   Kunnen, B. *et al.* Application of circularly polarised light for non-invasive diagnosis of cancerous tissues and turbid tissue-like scattering media. *Journal of Biophotonics* **8**, 317-323 (2015).

161    Xu, M. & Alfano, R. R. Circular polarisation memory of light. *Physical Review E* **72**, 065601 (2005).

162    Macdonald, C. M., Jacques, S. L., & Meglinski, I. V. Circular polarisation memory in polydisperse scattering media. *Physical Review E* **91**, 033204 (2015).

163    Wood, M. F. *et al.* Proof-of-principle demonstration of a Mueller matrix decomposition method for polarised light tissue characterisation in vivo. *Journal of Biomedical Optics* **14**, 014029 (2009).

164    Wang, Y. *et al.* Mueller matrix microscope: a quantitative tool to facilitate detections and fibrosis scorings of liver cirrhosis and cancer tissues. *Journal of Biomedical Optics* **21**, 71112 (2016).

165    Liu, T. *et al.* Distinguishing structural features between Crohn's disease and gastrointestinal luminal tuberculosis using Mueller matrix derived parameters. *Journal of Biophotonics* **12**, e201900151 (2019).

166    Shen, Y. *et al.* Comparative study of the influence of imaging resolution on linear retardance parameters derived from the Mueller matrix. *Biomedical Optics Express* **12**, 211-225 (2021).

167    Pierangelo, A. *et al.* Multispectral Mueller polarimetric imaging detecting residual cancer and cancer regression after neoadjuvant treatment for colorectal carcinomas. *Journal of Biomedical Optics* **18**, 046014 (2013).

168    Sun, M. *et al.* Characterising the microstructures of biological tissues using Mueller matrix and transformed polarisation parameters. *Biomedical Optics Express* **5**, 4223-4234 (2014).

169    Li, P., Lv, D., He, H. & Ma, H. Separating azimuthal orientation dependence in polarisation measurements of anisotropic media. *Optics Express* **26**, 3791-3800 (2018).

170    Gil, J. J. Invariant quantities of a Mueller matrix under rotation and retarder transformations. *Journal of Optical Society of America A* **33**, 52-58 (2016).

171    Iqbal, M., Ahmad, I., Khaliq, A. & Khan, S. Comparative study of Mueller matrix transformation and polar decomposition for optical characterisation of turbid media. *Optik* **224**, 165508 (2020).

172    Sun, T. *et al.* Distinguishing anisotropy orientations originated from scattering and birefringence of turbid media using Mueller matrix derived parameters. *Optics Letters* **43**: 4092-4095 (2018).

173    Khaliq, A. *et al.* Comparative study of 3 x 3 Mueller matrix transformation and polar decomposition. *Optics Communications* **485**: 126756 (2021).

174    Tariq, A. *et al.* Physically realizable space for the purity-depolarisation plane for polarised light scattering media. *Physical Review Letters* **119**: 033202 (2017).

175    Swami, M. *et al.* Polar decomposition of 3×3 Mueller matrix: a tool for quantitative tissue polarimetry. *Optics Express* **14**, 9324-9337 (2006).

176    Wang, Y. *et al.* Study on the validity of 3×3 Mueller matrix decomposition. *Journal of Biomedical Optics* **20**, 065003 (2015).

177    Morio, J. & Goudai, F. Influence of the order of diattenuator, retarder, and polariser in polar decomposition of Mueller matrices. *Optics Letters* **29**, 2234-2236 (2014).

178    Ghosh, N., Wood M. F., & Vitkin I. A. Influence of the order of the constituent basis matrices on the Mueller matrix decomposition-derived polarisation parameters in complex turbid media such as biological tissues. *Optics Communications* **283**, 1200-1208 (2010).

179    Li, P. *et al.* Analysis of tissue microstructure with mueller microscopy: logarithmic decomposition and monte carlo modeling. *Journal of Biomedical Optics* **25**, 015002 (2020).

180    Li, P., Tariq, A., He, H. & Ma, H. Characteristic Mueller matrices for direct assessment of the breaking of symmetries. *Optics Letters* **45**, 706-709 (2020).

181    Vizet, J. *et al.* In vivo imaging of uterine cervix with a Mueller polarimetric colposcope. *Scientific Reports* **7**, 2471 (2017).

182    Zhanghao, K. *et al.* High-dimensional super-resolution imaging reveals heterogeneity and dynamics of subcellular lipid membranes. *Nature Communications* **11**, 5890 (2020).

183    Zhanghao, K. *et al.* Super-resolution imaging of fluorescent dipoles via polarised structured illumination microscopy. *Nature Communications* **10**, 4694 (2019).

184    Spandana, K. U., Mahato, K. K. & Mazumder, N. Polarisation-resolved Stokes-Mueller imaging: a review of technology and applications. *Lasers in Medical Science* **34**, 1283-1293 (2019).

185    Yang, B. *et al.* Polarised light spatial frequency domain imaging for non-destructive quantification of soft tissue fibrous structures. *Biomedical Optics Express* **6**, 1520-1533 (2015).

186    Clancy, N. T. *et al.* Polarised stereo endoscope and narrowband detection for minimal access surgery. *Biomedical Optics Express* **5**, 4108-4117 (2014).

187    Manhas, S. *et al.* Demonstration of full 4×4 Mueller polarimetry through an optical fiber for endoscopic applications. *Optics Express* **23**, 3047-3054 (2015).

188  Wood, T. C. & Elson, D. S. Polarisation response measurement and simulation of rigid endoscopes. *Biomedical Optics Express* **1**, 463-470 (2010).

189  Vizet, J. *et al.* Optical fiber-based full Mueller polarimeter for endoscopic imaging using a two-wavelength simultaneous measurement method. *Journal of Biomedical Optics* **21**, 71106 (2016).

190  Rivet, S., Bradu, A., & Podoleanu, A. 70 kHz full 4x4 Mueller polarimeter and simultaneous fiber calibration for endoscopic applications. *Optics Express* **23**, 23768-23786 (2015).

191  Forward, Sarah, *et al.* Flexible polarimetric probe for 3× 3 Mueller matrix measurements of biological tissue. *Scientific Reports* **7**, 1-12 (2017).

192  Backman, V. *et al.* Polarised light scattering spectroscopy for quantitative measurement of epithelial cellular structures in situ. *IEEE Journal on Selected Topics in Quantum Electronics* **5**, 1019-1026 (1999).

193  Chan, D. *et al.* In vivo spectroscopic ellipsometry measurements on human skin. *Journal of Biomedical Optics* **12**, 014023 (2007).

194  Banerjee, P. *et al.* Probing the fractal pattern and organization of Bacillus thuringiensis bacteria colonies growing under different conditions using quantitative spectral light scattering polarimetry. *Journal of Biomedical Optics* **18**, 035003 (2013).

195  Soni, J. *et al.* Quantitative fluorescence and elastic scattering tissue polarimetry using an Eigenvalue calibrated spectroscopic Mueller matrix system. *Optics Express* **21**, 15475-15489 (2013).

196  Jagtap, J. *et al.* Quantitative Mueller matrix fluorescence spectroscopy for precancer detection. *Optics Letters* **39**, 243-246 (2014).

197  Satapathi, S., Soni, J. & Ghosh, N. Fluorescent Mueller matrix analysis of a highly scattering turbid media. *Applied Physics Letters* **104**, 131902 (2014).

198  Huse, N., Schönle, A. & Hell, S. W. Z-polarised confocal microscopy. *Journal of Biomedical Optics* **6**, 273-276 (2001).

199  Lim, N. S. J. *et al.* Early detection of biomolecular changes in disrupted porcine cartilage using polarised Raman spectroscopy. *Journal of Biomedical Optics* **16**, 017003 (2011).

200  Ahlawat, S. *et al.* Polarised Raman spectroscopic investigations on hemoglobin ordering in red blood cells. *Journal of Biomedical Optics* **19**, 087002 (2014).

201	Chan, K. H. *et al.* Use of 2D images of depth and integrated reflectivity to represent the severity of demineralization in cross-polarisation optical coherence tomography. *J Biophotonics* **8**, 36-45 (2015).

202	de Boer, J. F. & Milner, T. E. Review of polarisation sensitive optical coherence tomography and Stokes vector determination. *Journal of Biomedical Optics* **7**, 359-371, (2002).

203	Fan, C. M. & Yao, G. Imaging myocardial fiber orientation using polarisation sensitive optical coherence tomography. *Biomedical Optics Express* **4**, 460-465 (2013).

204	Gladkova, N. *et al.* Evaluation of oral mucosa collagen condition with cross-polarisation optical coherence tomography. *Journal of Biophotonics* **6**, 321-329 (2013).

205	Hitzenberger, C. *et al.* Measurement and imaging of birefringence and optic axis orientation by phase resolved polarisation sensitive optical coherence tomography. *Optics Express* **9**, 780-790 (2001).

206	Hong, Y. J., Makita, S., Sugiyama, S. & Yasuno, Y. Optically buffered Jones-matrix-based multifunctional optical coherence tomography with polarisation mode dispersion correction. *Biomedical Optics Express* **6**, 225-243 (2015).

207	Jiao, S., Yao, G. & Wang, L. V. Depth-resolved two-dimensional Stokes vectors of backscattered light and Mueller matrices of biological tissue measured with optical coherence tomography. *Applied Optics* **39**, 6318-6324 (2000).

208	Kuranov, R. V. *et al.* Complementary use of cross-polarisation and standard OCT for differential diagnosis of pathological tissues. *Optics Express* **10**, 707-713 (2002).

209	Lee, R. C., Kang, H., Darling, C. L. & Fried, D. Automated assessment of the remineralization of artificial enamel lesions with polarisation-sensitive optical coherence tomography. *Biomedical Optics Express* **5**, 2950-2962 (2014).

210	Popescu, D. P., Sowa, M. G., Hewko, M. D. & Choo-Smith, L. P. Assessment of early demineralization in teeth using the signal attenuation in optical coherence tomography images. *Journal of Biomedical Optics* **13**, 054053 (2008).

211	Sugita, M. *et al.* Retinal nerve fiber bundle tracing and analysis in human eye by polarisation sensitive OCT. *Biomedical Optics Express* **6**, 1030-1054 (2015).

212	Ugryumova, N., Attenburrow, D. P., Winlove, C. P. & Matcher, S. J. The collagen structure of equine articular cartilage, characterised using polarisation-sensitive optical coherence tomography. *Journal of Physics D Applied Physics* **38**, 2612-2619 (2005).

213	Vitkin, A., Ghosh, N. & de Martino, A. Tissue polarimetry. *Photonics: Scientific Foundations, Technology and Applications* **4**, 239-321 (2015).

214	Wang, L. & Zimnyakov, D. *Optical polarisation in biomedical applications*. (Springer, 2006).

215	Wang, L. H. V., Cote, G. L. & Jacques, S. L. Special section guest editorial - Tissue polarimetry. *Journal of Biomedical Optics* **7**, 278-278 (2002).

216	Wang, L. V. & Wu, H.-i. *Biomedical optics: principles and imaging*. (John Wiley & Sons, 2012).

217	Wang, S. & Larin, K. V. Optical coherence elastography for tissue characterisation: a review. *Journal of Biophotonics* **8**, 279-302 (2015).

218	Yamanari, M. *et al.* Scleral birefringence as measured by polarisation-sensitive optical coherence tomography and ocular biometric parameters of human eyes in vivo. *Biomedical Optics Express* **5**, 1391-1402 (2014).

219	Yao, G. & Wang, L. V. Two-dimensional depth-resolved Mueller matrix characterisation of biological tissue by optical coherence tomography. *Optics Letters* **24**, 537-539 (1999).

220	Milione, G. *et al.* Measuring the self-healing of the spatially inhomogeneous states of polarisation of vector Bessel beams. *Journal of Optics* **17**, 035617 (2015).

221	Mansfield, J. C., Winlove, C. P., Moger, J. & Matcher, S. J. Collagen fiber arrangement in normal and diseased cartilage studied by polarisation sensitive nonlinear microscopy. *Journal of Biomedical Optics* **13**, 044020 (2008).

222	Tanaka, Y. *et al.* Motion-artifact-robust, polarisation-resolved second-harmonic-generation microscopy based on rapid polarisation switching with electro-optic Pockells cell and its application to in vivo visualization of collagen fiber orientation in human facial skin. *Biomedical Optics Express* **5**, 1099-1113 (2014).

223	DeWalt, E. L. *et al.* Polarisation-modulated second harmonic generation ellipsometric microscopy at video rate. *Analytical Chemistry* **86**, 8448-8456 (2014).

224	Pavone, F. S. & Campagnola, P. J. *Second harmonic generation imaging*. (CRC Press, 2013).

225 Brasselet, S. Polarisation-resolved nonlinear microscopy: application to structural molecular and biological imaging. *Advances in Optics and Photonics* **3**, 205-271 (2011).

226 Kapsokalyvas, D. *et al.* In-vivo imaging of psoriatic lesions with polarisation multispectral dermoscopy and multiphoton microscopy. *Biomedical Optics Express* **5**, 2405-2419 (2014).

227 Golaraei, A. *et al.* Characterisation of collagen in non-small cell lung carcinoma with second harmonic polarisation microscopy. *Biomedical Optics Express* **5**, 3562-3567 (2014).

228 Daly, S. M. & Leahy, M. J. Go with the flow: A review of methods and advancements in blood flow imaging. *Journal of Biophotonics* **6**, 217-255 (2013).

229 Wolf, E. *Introduction to the Theory of Coherence and Polarisation of Light*. (Cambridge University Press, 2007).

230 Abrahamsson, S. *et al.* MultiFocus Polarisation Microscope (MF-PolScope) for 3D polarisation imaging of up to 25 focal planes simultaneously. *Optics Express* **23**, 7734-7754 (2015).

231 Chen, D. *et al.* Study of optical clearing in polarisation measurements by Monte Carlo simulations with anisotropic tissue-mimicking models. *Journal of Biomedical Optics* **21**, 081209 (2016).

232 Wang, Y. *et al.* Differentiating characteristic microstructural features of cancerous tissues using Mueller matrix microscope. *Micron* **79**, 8-15 (2015).

233 Hafi, N. *et al.* Fluorescence nanoscopy by polarisation modulation and polarisation angle narrowing. *Nature Methods* **11**, 579-584 (2014).

234 Cruz, C. A. V. *et al.* Quantitative nanoscale imaging of orientational order in biological filaments by polarised superresolution microscopy. *Proceedings of the National Academy of Sciences of the United States of America* **113**, E820-E828 (2016).

235 Kampmann, M., Atkinson, C. E., Mattheyses, A. L. & Simon, S. M. Mapping the orientation of nuclear pore proteins in living cells with polarised fluorescence microscopy. *Nature Structural & Molecular Biology* **8**, 643-649 (2011).

236 Sase, I., Miyata, H., Ishiwata, S. & Kinosita, K., Jr. Axial rotation of sliding actin filaments revealed by single-fluorophore imaging. *Proceedings of the National Academy of Sciences of the United States of America* **94**, 5646-5650, doi:10.1073/pnas.94.11.5646 (1997).

237	Forkey, J. N. *et al.* Three-dimensional structural dynamics of myosin V by single-molecule fluorescence polarisation. *Nature* **422**, 399-404 (2003).

238	Sosa, H., Peterman, E. J., Moerner, W. E. & Goldstein, L. S. ADP-induced rocking of the kinesin motor domain revealed by single-molecule fluorescence polarisation microscopy. *Nature Structural & Molecular Biology* **8**, 540-544 (2001).

239	DeMay, B. S. *et al.* Septin filaments exhibit a dynamic, paired organization that is conserved from yeast to mammals. *Journal of Cell Biology* **193**, 1065-1081 (2011).

240	DeMay, B. S., Noda, N., Gladfelter, A. S. & Oldenbourg, R. Rapid and quantitative imaging of excitation polarised fluorescence reveals ordered septin dynamics in live yeast. *Biophysical Journal* **101**, 985-994 (2011).

241	Axelrod, D. Carbocyanine dye orientation in red cell membrane studied by microscopic fluorescence polarisation. *Biophysical Journal* **26**, 557-573 (1979).

242	Schtz, G. J., Schindler, H. & Schmidt, T. Imaging single-molecule dichroism. *Optics Letters* **22**, 651-653 (1997).

243	Du, E. *et al.* Mueller matrix polarimetry for differentiating characteristic features of cancerous tissues. *Journal of Biomedical Optics* **19**, 76013 (2014).

244	Pierangelo, A. *et al.* Polarimetric imaging of uterine cervix: a case study. *Optics Express* **21**, 14120-14130 (2013).

245	Rehbinder, J. *et al.* Ex vivo Mueller polarimetric imaging of the uterine cervix: a first statistical evaluation. *Journal of Biomedical Optics* **21**, 71113 (2016).

246	Shukla, P. & Pradhan, A. Mueller decomposition images for cervical tissue: potential for discriminating normal and dysplastic states. *Optics Express* **17**, 1600-1609 (2009).

247	Chue-Sang, J. *et al.* Use of combined polarisation-sensitive optical coherence tomography and Mueller matrix imaging for the polarimetric characterisation of excised biological tissue. *Journal of Biomedical Optics* **21**, 71109 (2016).

248	Novikova, T. *et al.* The origins of polarimetric image contrast between healthy and cancerous human colon tissue. *Applied Physics Letters* **102**, 241103 (2013).

249     Ahmad, I. *et al.* Ex vivo characterisation of normal and adenocarcinoma colon samples by Mueller matrix polarimetry. *Journal of Biomedical Optics* **20**, 56012(2015).

250     Pierangelo, A. *et al.* Ex-vivo characterisation of human colon cancer by Mueller polarimetric imaging. *Optics Express* **19**, 1582-1593 (2011).

251     Pierangelo, A. *et al.* Ex vivo photometric and polarimetric multilayer characterisation of human healthy colon by multispectral Mueller imaging. *Journal of Biomedical Optics* **17**, 066009 (2012).

252     Dubreuil, M. *et al.* Mueller matrix polarimetry for improved liver fibrosis diagnosis. *Optics Letters* **37**, 1061-1063 (2012).

253     Wang, W. *et al.* Roles of linear and circular polarisation properties and effect of wavelength choice on differentiation between ex vivo normal and cancerous gastric samples. *Journal of Biomedical Optics* **19**, 046020 (2014).

254     Borovkova, M. *et al.* Evaluating beta-amyloidosis progression in Alzheimer's disease with Mueller polarimetry. *Biomedical Optics Express* **11**, 4509-4519 (2020).

255     Alali, S. *et al.* Assessment of local structural disorders of the bladder wall in partial bladder outlet obstruction using polarised light imaging. *Biomedical Optics Express* **5**, 621-629 (2014).

256     Mehta, S. B. *et al.* Dissection of molecular assembly dynamics by tracking orientation and position of single molecules in live cells. *Proceedings of the National Academy of Sciences of the United States of America* **113**, E6352-E6361 (2016).

257     Backman, V. *et al.* Detection of preinvasive cancer cells. *Nature* **406**, 35-36, (2000).

258     Qi, J. & Elson, D. S. A high definition Mueller polarimetric endoscope for tissue characterisation. *Scientific Reports* **6**, 26953 (2016).

259     Gan, Y. & Fleming, C. P. Extracting three-dimensional orientation and tractography of myofibers using optical coherence tomography. *Biomedical Optics Express* **4**, 2150-2165, (2013).

260     Pham, H. T. *et al.* Optical parameters of human blood plasma, collagen, and calfskin based on the Stokes-Mueller technique. *Applied Optics* **57**, 4353-4359 (2018).

261     Lu, R. W. *et al.* A polarisation-sensitive light field imager for multi-channel angular spectroscopy of light scattering in biological tissues. *Quantitative Imaging in Medicine and Surgery* **5**, 1-8 (2015).

262	Ghosh, N. *et al.* Mueller matrix decomposition for polarised light assessment of biological tissues. *Journal of Biophotonics* **2**, 145-156 (2009).

263	Wood, M. F. G. *et al.* Polarisation birefringence measurements for characterising the myocardium, including healthy, infarcted, and stem-cell-regenerated tissues. *Journal of Biomedical Optics* **15**, 047009 (2010).

264	Ahmad, I. *et al.* Polarimetric assessment of healthy and radiofrequency ablated porcine myocardial tissue. *Journal of Biophotonics* **9**, 750-759 (2016).

265	He, H. *et al.* Monitoring microstructural variations of fresh skeletal muscle tissues by Mueller matrix imaging. *Journal of Biophotonics* **10**, 664-673 (2017).

266	Sugita, S. & Matsumoto, T. Quantitative measurement of the distribution and alignment of collagen fibers in unfixed aortic tissues. *Journal of Biomechanics* **46**, 1403-1407 (2013).

267	Saytashev, I. *et al.* Self validating Mueller matrix Micro-Mesoscope (SAMMM) for the characterisation of biological media. *Optics Letters* **45**, 2168-2171 (2020).

268	Chen, Z. H., Meng, R. Y., Zhu, Y. H. & Ma, H. A collinear reflection Mueller matrix microscope for backscattering Mueller matrix imaging. *Optics and Lasers in Engineering* **129**, 106055 (2020).

269	Gonzalez, M. *et al.* Design and implementation of a portable colposcope Mueller matrix polarimeter. *Journal of Biomedical Optics* **25**, 116006 (2020).

270	Zotter, S. *et al.* Measuring retinal nerve fiber layer birefringence, retardation, and thickness using wide-dield, high-speed polarisation sensitive spectral domain OCT. *Investigative Ophthalmology & Visual Science* **54**, 72-84 (2013).

271	Dong, Y. *et al.* A quantitative and non-contact technique to characterise microstructural variations of skin tissues during photo-damaging process based on Mueller matrix polarimetry. *Scientific Reports* **7**, 14702 (2017).

272	Liu, X. *et al.* Tissue-like phantoms for quantitative birefringence imaging. *Biomedical Optics Express* **8**, 4454-4465 (2017).

273	Swami, M. K. *et al.* Effect of gold nanoparticles on depolarisation characteristics of Intralipid tissue phantom. *Optics Letters* **38**, 2855-2857 (2013).

274	Guo, Y. *et al.* Study on retardance due to well-ordered birefringent cylinders in anisotropic scattering media. *Journal of Biomedical Optics* **19**, 065001 (2014).

275   Li, X. *et al.* Polarimetric learning: a Siamese approach to learning distance metrics of algal Mueller matrix images. *Applied Optics* **57**, 3829-3837 (2018).

276   Heinrich, C. *et al.* Mueller polarimetric imaging of biological tissues: classification in a decision-theoretic framework. *Journal of Optical Society of America A* **35**, 2046-2057 (2018).

277   Wetzstein, G. *et al.* Inference in artificial intelligence with deep optics and photonics. *Nature* **588**, 39-47 (2020).

278   Rivenson, Y. *et al.* Deep learning microscopy. *Optica* **4**, 1437-1443 (2017).

279   He, C., Hu, Q., Dai, Y. & Booth, M. J. in *Adaptive Optics: Analysis, Methods & Systems.* OF2B. 5 (Optical Society of America).

280   Hu, Q., Dai, Y. Y., He, C. & Booth, M. J. Arbitrary vectorial state conversion using liquid crystal spatial light modulators. *Optics Communications* **459**, 125028 (2020).

281   Hu, Q., He, C. & Booth, M. J. Arbitrary complex retarders using a sequence of spatial light modulators as the basis for adaptive polarisation compensation. *Journal of Optics* (2021).

282   Dai, Y. Y. *et al.* Active compensation of extrinsic polarisation errors using adaptive optics. *Optics Express* **27**, 35797-35810 (2019).

283   Rubin, N. A. *et al.* Matrix Fourier optics enables a compact full-Stokes polarisation camera. *Science* **365**(6448), eaax1839 (2019).

284   Dorrah A. H. *et al.* Metasurface optics for on-demand polarisation transformations along the optical path. *Nature Photonics* **15**, 287-296 (2021).

285   Pan T. *et al.* Biophotonic probes for bio-detection and imaging. *Light: Science & Applications* **10**, 1-22 (2021).

286   Samim, M., Krouglov, S. & Barzda, V. Nonlinear Stokes-Mueller polarimetry. *Physical Review A* **93**, 013847 (2016).

287   Kontenis, L., Samim, M., Krouglov, S. & Barzda, V. in *CLEO: Science and Innovations.* STh4G. 6 (Optical Society of America).

288   Okoro, C. *Second-harmonic generation-based Mueller matrix polarisation analysis of collagen-rich tissues*, University of Illinois at Urbana-Champaign, (2018).

289   Krouglov, S. & Barzda, V. Three-dimensional nonlinear Stokes-Mueller polarimetry. *Journal of the Optical Society of America B-Optical Physics* **36**, 541-550, (2019).